\documentclass[
  journal=pasa,
  manuscript=research-paper, 
  year=2025,
  volume=YY,
]{cup-journal}

\usepackage{amsmath}
\usepackage{amssymb,microtype,siunitx,booktabs}
\usepackage[colorlinks=true, citecolor=blue, linkcolor=blue, urlcolor=blue]{hyperref}
\usepackage{multirow}

\title[PSR J1713+0747 Profile Change Event]{Ultra-Wideband Polarimetry of the April 2021 Profile Change Event in PSR J1713+0747}

\author{Rami~F. Mandow}
\affiliation{Department of Mathematics and Physical Sciences, Macquarie University, NSW 2109, Australia}
\alsoaffiliation{Australia Telescope National Facility, CSIRO, Space and Astronomy, P.O. Box 76, Epping, NSW 1710, Australia}
\email[Rami F. Mandow]{rami.mandow@hdr.mq.edu.au}

\author{Andrew Zic}
\affiliation{Australia Telescope National Facility, CSIRO, Space and Astronomy, P.O. Box 76, Epping, NSW 1710, Australia}
\alsoaffiliation{OzGrav: The ARC Center of Excellence for Gravitational Wave Discovery, Hawthorn VIC 3122, Australia}

\author{J.~R. Dawson}
\affiliation{Department of Mathematics and Physical Sciences, Macquarie University, NSW 2109, Australia}
\alsoaffiliation{Australia Telescope National Facility, CSIRO, Space and Astronomy, P.O. Box 76, Epping, NSW 1710, Australia}

\author{Shuangqiang Wang}
\affiliation{Xinjiang Astronomical Observatory, Chinese Academy of Sciences, Urumqi, Xinjiang 830011, People's Republic of China}
\alsoaffiliation{Australia Telescope National Facility, CSIRO, Space and Astronomy, P.O. Box 76, Epping, NSW 1710, Australia}

\author{Ma\l{}gorzata Cury\l{}o}
\affiliation{School of Physics and Astronomy, Monash University, Clayton VIC 3800, Australia}
\alsoaffiliation{OzGrav: The ARC Center of Excellence for Gravitational Wave Discovery, Clayton VIC 3800, Australia}

\author{Shi Dai}
\affiliation{Australia Telescope National Facility, CSIRO, Space and Astronomy, P.O. Box 76, Epping, NSW 1710, Australia}

\author{Valentina Di Marco}
\affiliation{School of Physics and Astronomy, Monash University, Clayton VIC 3800, Australia}
\alsoaffiliation{OzGrav: The ARC Center of Excellence for Gravitational Wave Discovery, Clayton VIC 3800, Australia}
\alsoaffiliation{Australia Telescope National Facility, CSIRO, Space and Astronomy, P.O. Box 76, Epping, NSW 1710, Australia}

\author{George Hobbs}
\affiliation{Australia Telescope National Facility, CSIRO, Space and Astronomy, P.O. Box 76, Epping, NSW 1710, Australia}

\author{Vivek Gupta}
\affiliation{Australia Telescope National Facility, CSIRO, Space and Astronomy, P.O. Box 76, Epping, NSW 1710, Australia}

\author{Agastya Kapur}
\affiliation{Department of Mathematics and Physical Sciences, Macquarie University, NSW 2109, Australia}
\alsoaffiliation{Australia Telescope National Facility, CSIRO, Space and Astronomy, P.O. Box 76, Epping, NSW 1710, Australia}
\alsoaffiliation{CSIRO Data61, Marsfield, NSW 2122, Australia}

\author{M.~Kerr}
\affiliation{Space Science Division, Naval Research Laboratory, Washington, DC 20375--5352, USA}

\author{Marcus~E. Lower}
\affiliation{Centre for Astrophysics and Supercomputing, Swinburne University of Technology, PO Box 218, Hawthorn, VIC 3122, Australia}

\author{Saurav Mishra}
\affiliation{Centre for Astrophysics and Supercomputing, Swinburne University of Technology, P.O. Box 218, Hawthorn, VIC 3122, Australia}
\alsoaffiliation{OzGrav: The ARC Center of Excellence for Gravitational Wave Discovery, Hawthorn VIC 3122, Australia}
\alsoaffiliation{Australia Telescope National Facility, CSIRO, Space and Astronomy, P.O. Box 76, Epping, NSW 1710, Australia}

\author{Daniel Reardon}
\affiliation{Centre for Astrophysics and Supercomputing, Swinburne University of Technology, P.O. Box 218, Hawthorn, VIC 3122, Australia}
\alsoaffiliation{OzGrav: The ARC Center of Excellence for Gravitational Wave Discovery, Hawthorn VIC 3122, Australia}

\author{Christopher~J. Russell}
\affiliation{CSIRO Scientific Computing, Australian Technology Park, Locked Bag 9013, Alexandria, NSW 1435, Australia}

\author{Ryan M. Shannon}
\affiliation{Centre for Astrophysics and Supercomputing, Swinburne University of Technology, P.O. Box 218, Hawthorn, VIC 3122, Australia}
\alsoaffiliation{OzGrav: The ARC Center of Excellence for Gravitational Wave Discovery, Hawthorn VIC 3122, Australia}

\author{Lei Zhang}
\affiliation{National Astronomical Observatories, Chinese Academy of Sciences, A20 Datun Road, Chaoyang District, Beijing 100101, People's Republic of China}
\alsoaffiliation{Centre for Astrophysics and Supercomputing, Swinburne University of Technology, P.O. Box 218, Hawthorn, VIC 3122, Australia}

\author{Xingjiang Zhu}
\affiliation{Department of Physics, Faculty of Arts and Sciences, Beijing Normal University, Zhuhai 519087, China}

\received {dd Mmm YYYY}
\revised  {dd Mmm YYYY}
\accepted {dd Mmm YYYY}
\published{22 September 202X}


\begin{document}

\begin{abstract}
The millisecond pulsar PSR~J1713$+$0747 is a high-priority target for pulsar timing array experiments due to its long-term timing stability, and bright, narrow pulse profile. In April 2021, PSR~J1713$+$0747 underwent a significant profile change event, observed by several telescopes worldwide. Using the broad bandwidth and polarimetric fidelity of the Ultra-Wideband Low-frequency receiver on Murriyang, CSIRO's Parkes radio telescope, we investigated the long-term spectro-polarimetric behaviour of this profile change in detail. We highlight the broad-bandwidth nature of the event, which exhibits frequency dependence that is inconsistent with cold-plasma propagation effects. We also find that spectral and temporal variations are stronger in one of the orthogonal polarisation modes than the other, and observe mild variations ($\sim 3$ -- $5\,\sigma$ significance) in circular polarisation above 1400\,MHz following the event. However, the linear polarisation position angle remained remarkably stable in the profile leading edge throughout the event. With over three years of data post-event, we find that the profile has not yet recovered back to its original state, indicating a long-term asymptotic recovery, or a potential reconfiguration of the pulsar's magnetic field. These findings favour a magnetospheric origin of the profile change event over a line-of-sight propagation effect in the interstellar medium.
\end{abstract}

\section{Introduction}
\label{sec:int}
Millisecond pulsars (MSPs - \citealt{1982Natur.300..615B,1982Natur.300..728A}) are extraordinarily stable rotators \citep{2009MNRAS.400..951V}, comparable to terrestrial atomic clocks over years to decades \citep{1991IEEEP..79.1054T}. While a large fraction of canonical pulsars exhibit large rotational irregularities \citep{2010MNRAS.402.1027H, 2025MNRAS.538.3104L} or frequent glitching \citep{2022MNRAS.510.4049B}, these phenomena are much more rare, or lower in magnitude, in MSPs \citep{2010ApJ...725.1607S,2016MNRAS.461.2809M}.

Time-integrated MSP pulse profiles have in general proven to be exceptionally stable compared with the canonical pulsar population \citep{2012MNRAS.420..361L}, reflecting the stability of their magnetospheres. The rotational and profile stability of MSPs can be exploited to carry out high-precision pulsar timing experiments, including precise pulsar mass measurements \citep{2024ApJ...971L..20C,2024ApJ...971L..18R,2024ApJ...971L..19R}, stringent tests of General Relativity \citep{2021PhRvX..11d1050K,2021MNRAS.504.2094K} and searches for gravitational waves \citep{2023ApJ...951L...8A,2023A&A...678A..50E,2023ApJ...951L...6R,2023RAA....23g5024X,2025MNRAS.536.1489M}.\\ 
\indent Despite the apparent stability of MSP integrated profiles, some MSPs have exhibited similar temporal profile variations to those seen in canonical pulsars. One class of pulsar profile variability is ``jitter'' \citep{2021MNRAS.502..407P}, which refers to variations of pulsar profile shapes on the timescale of individual rotations. This may lead to small variations in integrated pulse profiles, particularly for short observations. Short-term discrete profile changes, such as mode-changing and nulling \citep{2022MNRAS.510.5908M,2023MNRAS.523.4405N}, can also occur on short timescales, ranging from individual rotations to days. Stochastic profile variability on longer timescales of months -- years has been observed in both canonical pulsars \citep{2016MNRAS.456.1374B,2022MNRAS.513.5861S,2023MNRAS.524.5904L,2024MNRAS.528.7458B,2025MNRAS.538.3104L} and MSPs \citep{ 1997AJ....114.1539B,2011MNRAS.418.1258O,2018ApJ...868..122B,2021MNRAS.500.1178P,2023MNRAS.523.4405N,2025ApJ...984..139F}, and is thought to arise due to long-term variability in the pulsar magnetosphere.
Additionally, propagation effects through the interstellar medium, such as scattering \citep{2015MNRAS.449.1570L}, diffractive scintillation \citep{2022A&A...664A.116L}, and plasma lensing \citep{2011MNRAS.410..499G}, can cause profile variability, although these are clearly not intrinsic to the pulsar itself.

Discrete profile change events -- the subject of this paper -- manifest as a sudden augmentation of pulse profile components or the appearance of new components that decay over time. There are several noteworthy examples of long-term discrete profile changes reported in the MSP population: PSR J0437$-$4715 \citep{2021MNRAS.502..478G}, PSR J1643$-$1224 \citep{2016ApJ...828L...1S,2021MNRAS.502..478G}, and two prior events on PSR J1713$+$0747 (hereafter ``J1713") which we describe below. Transient profile changes have also been observed from several canonical pulsars \citep{2010Sci...329..408L,2011MNRAS.415..251K,2014ApJ...780L..31B,2016MNRAS.456.1374B,2022MNRAS.513.5861S,2025arXiv250523413K}. However, most of these transient profile events are different in character to the abrupt change followed by gradual decays seen in MSPs like J1713 -- for example, they may arise more gradually, manifest as one component of a long-term ``state-switching'' phenomenon, or do not exhibit a gradual decay, instead forming part of a continuous process of profile variability. Close analogues to the abrupt profile change event like PSR J1713$+$0747 have been observed in canonical pulsars like PSRs J0738$-$4042 \citep{2011MNRAS.415..251K, 2014ApJ...780L..31B}, B2035$+$36 \citep{2022MNRAS.513.5861S, 2025arXiv250523413K}, and J1119$-$6127, which showed erratic pulse shape changes shortly after large glitches \citep{2011MNRAS.411.1917W} and X-ray outbursts \citep{2018MNRAS.480.3584D}. Given the apparent long-term stability of most MSP magnetospheres, studies of this class of profile change event in MSPs may yield important clues into how the phenomenon occurs across the population as a whole, and to the emission mechanism itself. 

PSR~J1713$+$0747 \citep{1993ApJ...410L..91F} is a 4.57-millisecond-period MSP in a binary system with a white dwarf companion. Its long-term timing stability, bright, narrow pulse profile, and equatorial sky position has made it a centrepiece of many PTA observing programmes \citep{2023ApJ...951L...9A,2023A&A...678A..48E,2023PASA...40...49Z,2023RAA....23g5024X,2025MNRAS.536.1467M}

J1713 has undergone two frequency-dependent fluctuations in its times of arrival which have been attributed to profile change events. The first, in October 2008, lasted between 100 and 200 days before recovering \citep{2015ApJ...808..113C,2015ApJ...809...41Z,2016MNRAS.458.3341D}. The second event occurred $\sim$8 years later, in April 2016 \citep{2018ApJ...861..132L}. \cite{2021MNRAS.508.1115L} suggested both events likely originated from discrete dispersion measure (DM) variations. On the other hand, \cite{2021MNRAS.502..478G} concluded that the 2016 event was likely magnetospheric in origin.
Although there is no reported evidence for pulse shape changes during the 2008 event, \cite{2024ApJ...964..179J} proposed that both historical events were likely related to the magnetosphere.

In mid-April 2021, J1713 exhibited a significant change to its radio pulse profile \citep{2021ATel14642....1X,2021ATel14652....1M,2021ATel14667....1S}. The time of the profile change was determined to be between Modified Julian Day (MJD) 59320 -- 59321 \citep{2021ATel14652....1M}. Studies of the April 2021 event by \citet{2021MNRAS.507L..57S}, \citet{2024ApJ...964..179J} and  \citet{2024ApJ...961...48W} have revealed complex, frequency-dependent profile shape changes, suggesting a magnetospheric origin as opposed to propagation effects. These studies observed the profile recovery over time, reporting different timelines for its return to a near pre-event state \citep{2021MNRAS.507L..57S,2024ApJ...961...48W,2024ApJ...964..179J}. Both \cite{2021MNRAS.507L..57S} and \cite{2024ApJ...961...48W} also reported that higher frequencies exhibited more pronounced changes. These studies collectively disfavour alternative explanations such as glitches, dispersion measure changes, scintillation and jitter \citep{2021RNAAS...5..167L,2024ApJ...964..179J}.

Radio emission from pulsars is highly polarised, exhibiting frequency-dependent linear and circular polarisation profiles that have distinct morphological features \citep{1968Natur.218..124L,1998MNRAS.301..235G}. The polarisation profiles encode crucial information about the emission mechanism and propagation of radiation through the pulsar magnetosphere \citep{1999ApJ...526..957K,2011MNRAS.414.2087Y,2024MNRAS.532.3558K,2025A&A...695A.173X}. Linear polarisation is particularly important, as the observed polarisation position angle (PA) can be used to infer the magnetic field geometry of the pulsar via the `rotating vector model' \citep{1969Natur.221..443R}. While many pulsars show smooth PA tracks, some pulsars can also exhibit abrupt 90* jumps across pulse phase, corresponding to transitions between the orthogonal polarisation modes \citep[OPMs;][]{1975PASA....2..334M}. 

To date, no detailed study of the polarimetric behaviour of the J1713 profile change has been conducted. Discrete changes in polarisation properties, such as those induced by profile change events, therefore offer important insights into evolving magnetospheric conditions and sources of instability that may affect high-precision timing experiments. Wideband spectro-polarimetric observations are particularly crucial for probing these phenomena, offering access to information about the magnetosphere that would otherwise be inaccessible \citep{2020MNRAS.496.1418O,2023MNRAS.520.4961O,2024NatAs...8..606L}. Furthermore, because J1713 is one of the most important MSPs in global PTA campaigns, detailed characterisation of this profile change event is critical for informing mitigation strategies in precision timing analysis such as searches for low-frequency gravitational waves.

 In this paper, we present a wideband polarimetric analysis of the ongoing recovery of the profile of J1713 following the April 2021 change event -- the most significant profile change yet recorded on an MSP. Under the Parkes Pulsar Timing Array observing programme, J1713 has been regularly observed using the Ultra-Wideband Low-frequency receiver \citep[UWL; ][]{2020PASA...37...12H} on \textit{Murriyang}, CSIRO's Parkes Radio Telescope. With a frequency band spanning 704--4032\,MHz, well-calibrated polarimetry, and a long-term continuous wide-band dataset spanning $> 3$\,yr, this dataset offers the opportunity to characterise this profile change in greater depth than previous studies. 
 Observation details and methods are presented in Section \ref{sec:obsmethods}, analysis and findings in Section \ref{sec:results}, and further discussions in Section \ref{sec:discussion}.

\section{Observations and Methods}
\label{sec:obsmethods}
\subsection{Observations and Data Reduction}
Our observations were taken under two separate observing programmes on Murriyang: The Parkes Pulsar Timing Array programme \citep[PPTA; ][project code P456]{2013PASA...30...17M}, and a separate, targeted programme (P1172). The PPTA observes a set of 37 MSPs with approximately 3-week cadence (see \citealt{2023PASA...40...49Z} for details of the observing programme). Under the P1172 programme, we took targeted high-cadence ($\sim$2--7 days) observations of J1713 to more closely monitor the recovery from the April 2021 profile change event, between October 2022 to April 2024. All observations were carried out using the UWL receiver, recorded in pulsar fold mode and coherently de-dispersed to the pulsar DM value $15.988$\,pc\,cm$^{-3}$. The observations were typically 64\,minutes in duration, although occasionally observations were cut short for operational or instrumental reasons.

The observations were flagged, calibrated, and reduced following the same procedures used for the PPTA Third Data Release (PPTA DR3), as described in \citet{2023PASA...40...49Z}. In this work, we make use of the sub-banded data products, using the same frequency sub-band divisions defined in the PPTA DR3, which we list in Table \ref{tab:uwlfreq} for clarity. We installed the timing ephemeris from the PPTA DR3 onto all observation archive files. To ensure sufficient signal-to-noise ratio (S/N), we frequency-averaged the profiles within each of the 8 sub-bands. For each sub-band, we removed any observations with a S/N less than 50. We also excluded data with corrupted polarisation measurements due to digitiser synchronisation errors; this affected small frequency ranges for short amounts of time, most significantly around December 2021 -- February 2022 across 1.0 -- 1.3\,GHz and February 2022 -- May 2022 from 3.0 -- 3.5\,GHz.

\begin{table}
\centering
\caption{\label{tab:uwlfreq}
UWL sub-band ranges (where $\nu_\mathrm{min}$ and $\nu_\mathrm{max}$ represent the minimum and maximum frequency), bandwidth and centre frequency ($\nu_{c})$}
    \begin{tabular}{ccccc}
    \hline
    \hline
    \textbf{Sub-band} & \textbf{$\nu_{\text{min}}$ (MHz)} & \textbf{$\nu_{\text{max}}$ (MHz)} & \textbf{BW} (MHz) & \textbf{$\nu_{\text{c}}$}\\
    \hline
    sbA & 704 & 768 & 64 & 736\\ 
    sbB & 768 & 864 & 96 & 816\\ 
    sbC & 864 & 960 & 96 & 912\\ 
    sbD & 960 & 1248 & 288 & 1104\\ 
    sbE & 1248 & 1504 & 256 & 1376\\ 
    sbF & 1504 & 2048 & 544 & 1776\\ 
    sbG & 2048 & 2592 & 544 & 2320\\ 
    sbH & 2592 & 4032 & 1440 & 3312\\
    \hline 
    \end{tabular}
\end{table}

In this work, we analyse changes to linear polarisation PAs over time. This requires measurement and correction for variations in the interstellar and ionospheric rotation measure (RM), in addition to the data processing performed in the PPTA DR3. To this end, we used the \textsc{rmfit} tool in the \textsc{psrchive} software suite \citep{2004PASA...21..302H,2011PASA...28....1V} to measure the RM for each observation, manually setting the RM bounds from -10 to 30 $\text{rad m}^{-2}$, and using 40 steps. We then used the general-purpose pulsar data manipulation tool \textsc{pam} to correct the profile archives for RM at an infinite reference frequency, using the \textsc{-\phantom{}-aux\_rm} option.

\begin{figure*}
    \centering
    \includegraphics[height=0.9\textheight]{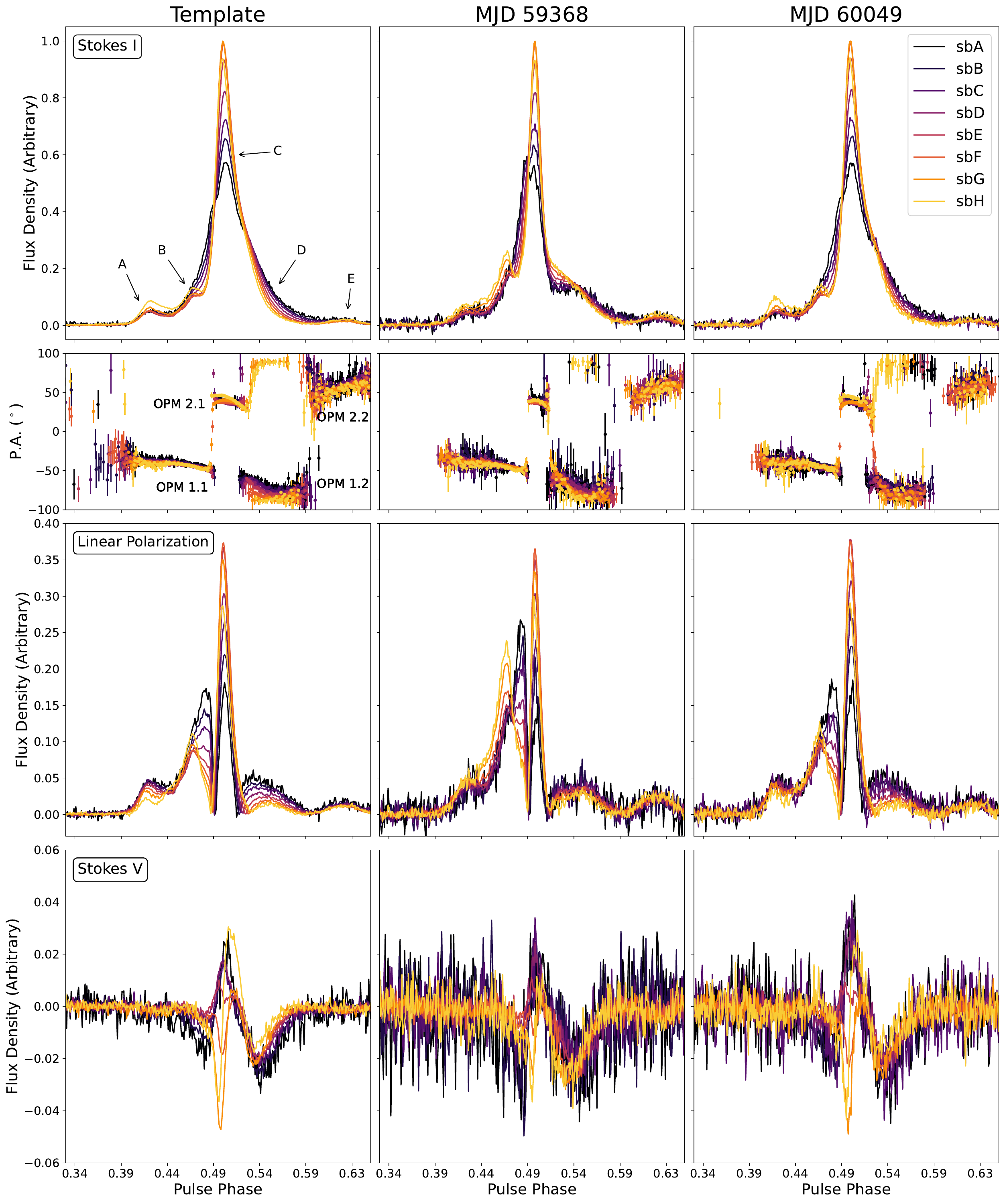}
    \caption{Stokes $I$, PA, Linear polarisation and Stokes $V$ profiles colour-mapped as a function of frequency per row. The first column represents the template profiles, the second column represents MJD 59368 (47 days post-event) and the last column represents MJD 60049 (roughly two years post-event). Annotated are five profile components: (A) the leading peak, (B) the profile shoulder, (C) the main pulse peak, (D) the descending gradient, and (E) the trailing peak. The four OPM sub-components are also annotated. In addition to the normalisation by the integrated Stokes $I$ flux density, we additionally normalise the Stokes $I$, $L$, and Stokes $V$ profiles for each epoch in this plot by the maximum of normalised Stokes $I$ flux densities across all sub-bands, scaling the normalised intensities to a maximum of 1.}
    \label{IPALV_3epoch}
\end{figure*}

\subsection{Profile alignment}
The dramatic and sustained change in the profile shape of J1713 (see Figures \ref{IPALV_3epoch},  \ref{fig:waterfalls_Eph}) means that one of the fundamental assumptions of pulsar timing -- that the pulse shape remains stable over time -- is violated. This leads to a degeneracy between profile variability (e.g. due to magnetospheric changes) and pulse arrival times, because a mismatch between a reference template and the true profile can be erroneously modeled as a profile phase shift. Conversely, if there were variations to the rotational properties of the pulsar, the change in profile shape makes it challenging to disentangle rotational effects from those that purely augment the observed pulse profile (e.g. in the pulsar magnetosphere). For the pulse profile analyses presented in this paper, one possible assumption is that the intrinsic timing properties of the pulsar have not changed. In this case, the expected rotational phase location of the radio pulse is determined by the ephemeris prediction. While this is subject to usual variations that affect pulsar timing (e.g. DM variations, timing noise, secular changes in orbital parameters, etc.), using an established timing ephemeris to predict rotational phase allows careful comparison of specific sub-components of the pulse profile which may (or may not) have changed significantly. On the other hand, if we relax the assumption that the intrinsic rotational properties did not change, the profile change can be studied by minimising the difference between the observed pulse profile and a standard template. This necessitates aligning the observed profiles with this standard template, and, by design, minimises the difference between the observed profile and the template. In practice, we achieve this by using the \citet{1992PTRSL.341..117T} Fourier phase-gradient alignment algorithm, fitting the phase shift of the profiles using the templates which we describe in Section \ref{subsec:template_create}. A genuine change to the intrinsic rotation wouldn't produce the profile residuals that we observe in the FFT-aligned methodoly we have applied, and the appearance of such residuals are the result of intrinsic profile shape changes. In the analyses presented here, we employ both the ephemeris-predicted, and template-aligned methods to predict the rotational phase location of the pulse profiles over time. 

\subsection{Sub-banded Template Generation}
\label{subsec:template_create}
We used all available UWL observations prior to the event date to construct frequency-averaged templates of all four polarisations ($I$, $Q$, $U$, $V$) for each of the constructed sub-bands. These observations spanned from October 2018 to the end of March 2021.

The polarised, sub-band profile templates were generated as follows. We first estimate the total intensity S/N in every pre-event observation using \textsc{psrstat}, and sort these profiles in descending order according to their S/N. We then begin an iterative process of constructing the templates, beginning with the highest S/N profile as the initial reference. We use the \citet{1992PTRSL.341..117T} Fourier phase-gradient alignment algorithm to align the second-highest S/N profile to the first. This phase shift is computed on the total intensity profile but applied to all four Stokes parameters ($I$, $Q$, $U$, $V$). We then construct a new template by taking the weighted mean of all of the aligned profiles, which is then used as the template profile in the next step. We apply this process iteratively until all profiles are aligned. The template thus generated, is the mean of all aligned profiles. 

This template is used to produce profile residuals both for the ephemeris-predicted and template-aligned profiles. For the ephemeris-predicted dataset, we adjust the absolute phase of the template by taking a simple un-aligned mean of all pre-event profiles, calculating the phase offset between our constructed template and this simple mean, and shifting the template by this offset. This ensures that the profile residuals on the ephemeris-predicted profiles are not biased by any timing deviations present in the highest S/N profile. 

We construct the template and sample profiles for the total linearly polarised intensity (hereafter $L$) using the polarised intensity image-based debiasing technique detailed in \cite{2017A&A...600A..63M}, modified to operate on our one-dimensional profiles. This is used to mitigate the Ricean noise statistics that bias na\"ive calculations of $L = \sqrt{Q^2 + U^2}$ when using noisy data. This method provides superior performance at low profile S/N compared to the method of \citet{2001ApJ...553..341E} commonly used in the pulsar literature, which truncates $L$ profiles below a S/N of 1.57. The  \cite{2017A&A...600A..63M} debiasing method provides a continuous estimate of $L$ regardless of the profile S/N,  while also preserving the zero-mean gaussian noise statistics of the input Stokes $Q$ and $U$ profiles in the final $L$ profile. The linear polarization position angle of the template, and sample profiles are computed as $\text{PA} = 1/2~\text{arctan}2(U,Q)$, where $\text{arctan}2$ is the two-argument arctangent.

We note that accurately modelling the frequency evolution of MSP profiles is challenging, and we do not specifically attempt to do so in this study. However, the stability of the pre-event residuals across all sub-bands over 2.4 years prior to the event (see Figure \ref{fig:waterfalls_Eph}), demonstrates that our sub-banded templates, constructed from long-term average profiles, are reliable and fit-for-purpose.

\subsection{Normalisation and construction of profile residuals}
Processes such as scintillation and intrinsic flux density modulation can induce frequency-dependent variability in pulsar profiles, including that of J1713 \citep{2002ApJ...581..495B, 2018ApJ...868..122B}. To compare pulse profiles across time, we therefore first normalise the template and observed profiles using the integrated flux density under the on-pulse region of its Stokes $I$ profile. We use a fixed on-pulse window over pulse phase bins 0.38 -- 0.68 to mitigate potential bias arising from the S/N dependence in determining the on-pulse window. The normalisation procedure ensures that variations in the profile shape, rather than its amplitude, become the primary focus of our analysis.  

In some epochs we found a small flux density offset between the peak of the normalised template and normalised observation profile, particularly in low S/N epochs, which introduced a small artificial residual in the pre-event data. We corrected this by performing a least-squares fit to determine the optimal scale factor between the aligned profile and template within the central pulse component (phase 0.462--0.562). These scale factors were then stored and also applied to the ephemeris-predicted profiles, ensuring consistent flux density scaling and minimising normalisation-induced artifacts in the residuals. 

We constructed the profile residuals by subtracting the Stokes $I$, Stokes $V$, or $L$ template from the corresponding observation. We carefully validated the data by inspecting each profile and its residuals by eye, and produced histograms of the off-pulse regions of the profile to identify observations corrupted by excessive interference or instrumental issues. Any suspect data were inspected, then accepted, reprocessed or rejected as applicable.

\section{Results}
\label{sec:results}

\begin{figure*}
    \centering
    \includegraphics[width=18cm]{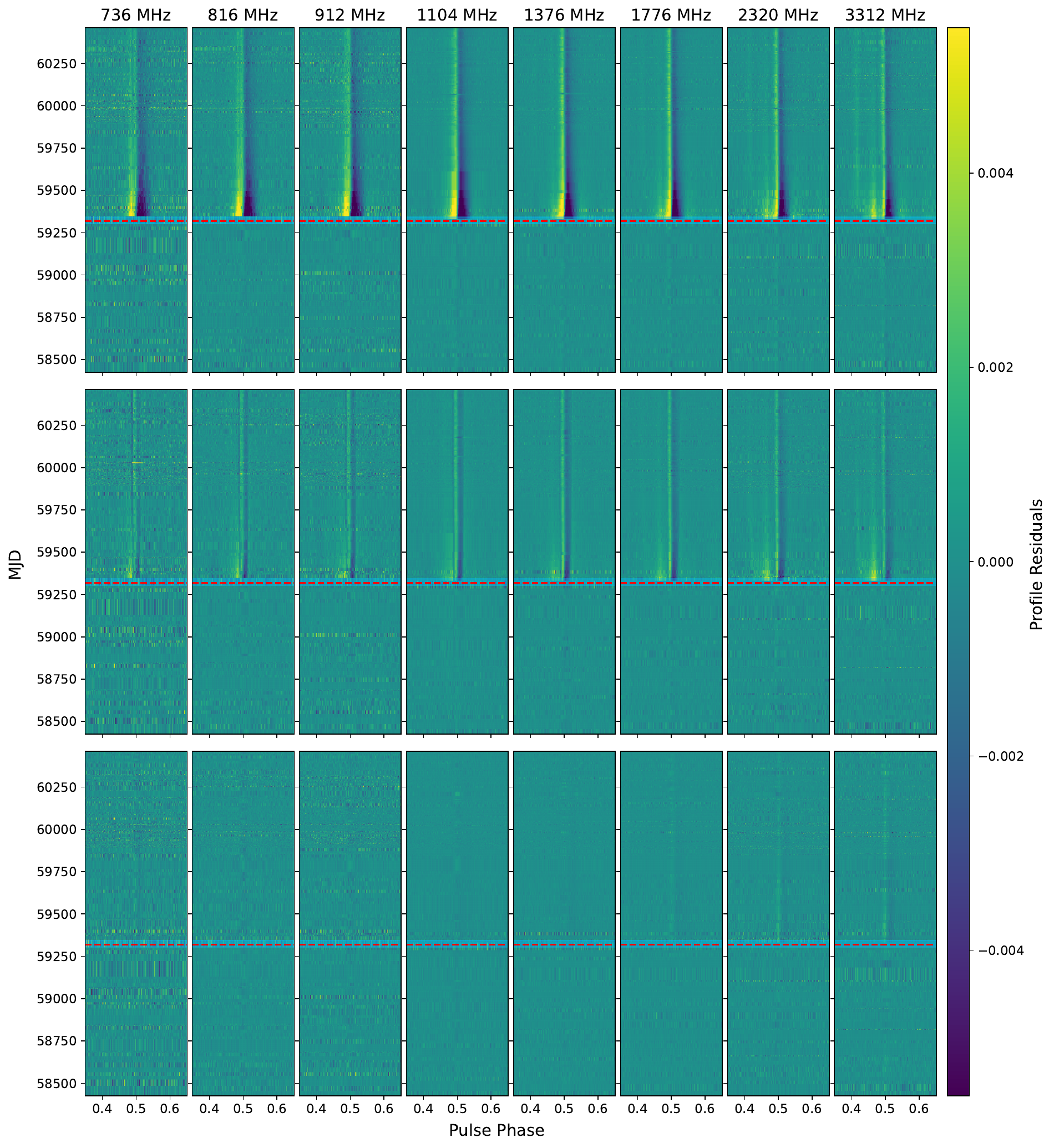}
    \caption{Stokes $I$, $L$ and Stokes $V$ profile residuals as a function of time, where pulse phase at each epoch is predicted from the timing ephemeris. The top row shows Stokes $I$, the middle row shows linearly polarised intensity (``$L$''), and the bottom row shows Stokes $V$. The red dashed line indicates the time of the profile change event. The colour bar indicates the intensity of the profile residual. As can be seen, Stokes $I$ and $L$ were affected, whereas Stokes $V$ was minimally affected in the lower sub-bands A--E, but shows variations in the higher-frequency sub-bands sbF -- sbH. Additionally, the profile has not yet returned to its pre-event state, indicating either a long recovery timescale or potential reconfiguration of the profile.}
    \label{fig:waterfalls_Eph}
\end{figure*}

\begin{figure*}
    \centering
    \includegraphics[width=18cm]{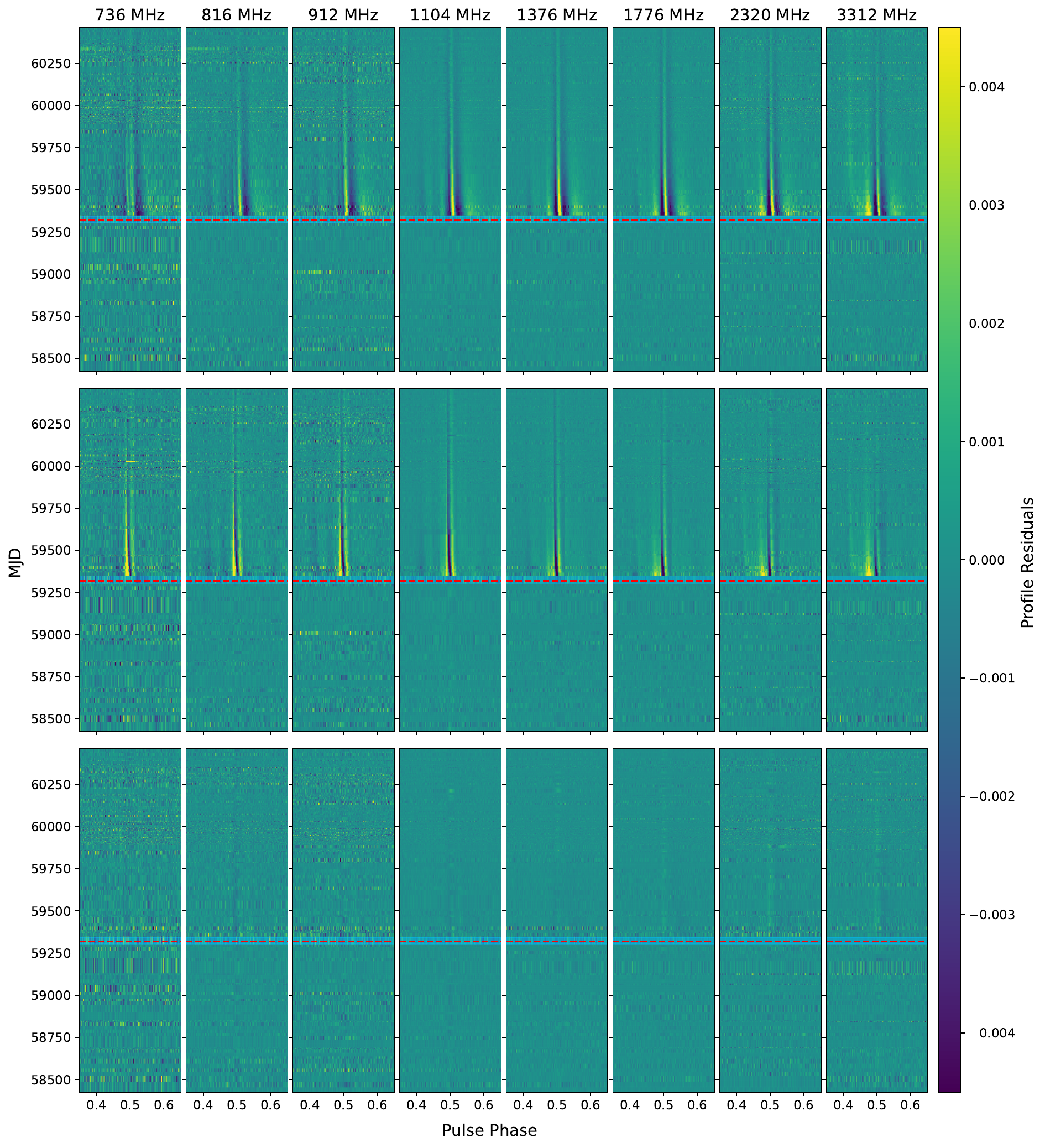}
    \caption{As in Figure \ref{fig:waterfalls_Eph}, but for the case where the profiles have been aligned to the template.}
    \label{fig:waterfalls_FFT}
\end{figure*}

In this section, we present our results, starting with an analysis of profile shape changes for the different Stokes parameters, followed by a Principal Component Analysis of the profile residuals. We then examine variability in the linear polarisation, position angle, and orthogonal polarisation modes.

\subsection{Profile shape variability}
\label{ProVar}
In Figure~\ref{IPALV_3epoch} we present the template for J1713 (first column) as well as two sample high S/N profiles that highlight the profile change effects at two epochs -- MJD 59368 which is 47 days after the event, and MJD 60049, approximately two years after the event. Also included are the PA data. The frequency across the UWL spanning the eight sub-bands is colour-mapped with lower frequencies as the darker curves and higher frequencies as lighter curves, and Stokes $I$, PA, $L$, and Stokes $V$ are plotted on separate rows. Annotated are five profile components, labelled: (A) the leading peak, (B) the profile shoulder, (C) the main pulse peak, (D) the descending gradient, and (E) the trailing peak. At early times after the profile change event, the central peak of the Stokes $I$ profile (profile component C) significantly narrows across the entire band, whilst profile component D broadens into a plateau across pulse phase -- increasing the width of the profile base and forming a distinct horizontal curve before resuming its decline. This horizontal feature is visible across all sub-bands, except sbG and sbH. Above approximately 1500\,MHz, the leading shoulder (profile component B at pulse phase bins $\sim$0.45--0.50) intensifies with increasing frequency.  

The linear polarisation template exhibits two main peaks close to the centre, with two smaller trailing peaks. Each of these structures corresponds to a different OPM sub-component (see Section \ref{sec:OPM_changes} for further details). These different OPM sub-components have been labelled in the first PA sub-plot. As Figure \ref{IPALV_3epoch} highlights, there were dramatic, complex frequency-dependent changes in the relative heights of the linear polarisation profile components. These correlate with changes in the PA -- with sub-component 2.1 narrowing, and sub-component 1.2 forming modal bridges to its neighbouring component. Further details of these variations are discussed in Section \ref{sec:OPM_changes_PA}. A prominent change in the first peak of the linear polarisation profile is also observed, where the peak height increases dramatically in a non-monotonic frequency-dependent manner. As shown in the central sub-plot for linear polarisation of Figure Section \ref{IPALV_3epoch}, which represents the epoch soon after the event, the higher-frequency bands exhibit an increase by a factor of $\sim$2.5 compared to pre-event levels, before gradually returning to baseline values. This first peak corresponds to the OPM sub-component 1.1 (see Section \ref{sec:OPM_changes_opmflux} for further details).

While minor variations are seen in Stokes $V$ in this figure, the overall flux density and scale of change remain much smaller than Stokes $I$ and linear polarisation.

Figures \ref{fig:waterfalls_Eph} and \ref{fig:waterfalls_FFT} show the evolution of the profile residuals over time, spanning from the beginning of our UWL observations in October 2018 (MJD 58426) and extending to the start of June 2024 (MJD 60462). Figure  \ref{fig:waterfalls_Eph} shows the profile residuals from ephemeris-predicted profiles, and Figure \ref{fig:waterfalls_FFT} shows the template-aligned profiles. 
In both sets of residuals, we observe that Stokes $I$ and linear polarisation exhibit significant changes compared to the pre-event data across the full band. In contrast, Stokes $V$, which has the lowest flux density and therefore lowest S/N, does not show a  significant deviation post-event at frequencies below $\sim$1400\,MHz (sub-bands A -- E), with fewer than 4\% of post-event observations having peak residual values above $3\sigma$ significance. On the other hand, inspection of the post-event residuals reveals significant profile variations at higher frequencies, as can be seen in Figures \ref{fig:waterfalls_Eph} and \ref{fig:waterfalls_FFT}, with 13--21\% of observations showing residual peaks above $3\sigma$ significance, across sbF -- sbH (1376\,MHz -- 3312\,MHz). 

Additionally, new profile features emerge across the entire band. While most of these features gradually diminish over time, some -- particularly in the higher frequency bands ($\nu_{c}$: 2320\,MHz; 3312\,MHz) -- persist. Notably, a stable positive residual feature at approximately pulse phase 0.42 in the highest sub-band remains present through to our final observation, indicating the emergence of a new structure in the pulse profile that persists over long timescales. The overall changes in the linear polarisation profile residuals are narrower in pulse phase compared to those in Stokes $I$, yet still span the entire UWL bandwidth. 

The residuals from our template-aligned profiles (Figure \ref{fig:waterfalls_FFT}) show consistent results with the residuals from ephemeris-predicted profiles. New profile features are also observed under this alignment method across the band, gradually diminishing over time. The positive feature at pulse phase 0.42 in the highest sub-bands remains evident in this dataset. Using this alignment method, we also observe a sudden right-ward jump in phase followed by a gradual left-ward drift in the central components for approximately one year following the event, which is not observed in the ephemeris-aligned residuals. This is a non-physical artifact resulting from the template alignment process, as this method is sensitive to time-dependent shape changes. Due to this introduced bias, we consider the template-alignment method sub-optimal relative to the ephemeris-aligned method.

\subsubsection{Principal Component Analysis on Profile Variability}
\label{PolVar}
Following \citet{2024ApJ...964..179J}, we performed a principal component analysis (PCA) on both the Stokes $I$ and linear polarisation residuals (using the \textsc{sklearn} library, \citealt{scikit-learn}), in order to quantify and model the ongoing recovery of the profile. In essence, this technique quantifies the complex pulse phase-dependent variations using a small number of principal components. It works by decomposing the covariance matrix of the profile residuals into orthogonal principal components (eigenvectors) that are ranked by the degree of the residual structure they explain (i.e. ranked by the magnitude of the corresponding eigenvalues). We conducted PCA on each individual sub-band, yielding PC scores that quantify the magnitude of each of the principal components over time.

\begin{figure*}
    \centering
    \includegraphics[width=\linewidth]{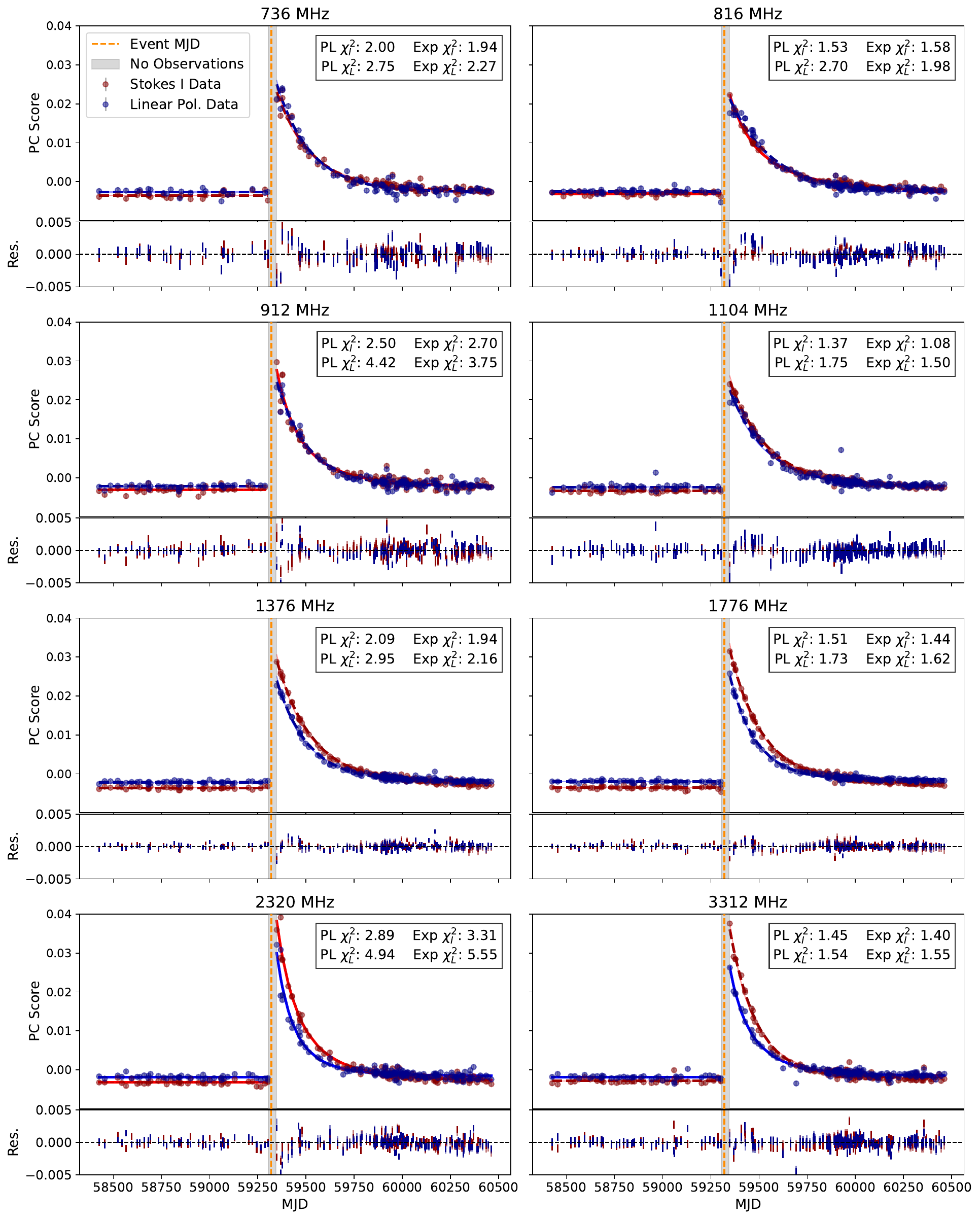}
    \caption{Results of PCA analysis showing the first principal component scores in Stokes $I$(red) and linear polarisation (blue) as a function of time. The top subpanel of each plot shows the fit to the PC score data, and the bottom subpanel shows the fit residuals. A solid line indicates that the power-law model is preferred, where a dashed line indicates the exponential model. The orange dashed line indicates the profile change event, and the surrounding shaded grey region indicates the period of no observations. The $\chi^{2}_{\mathrm{r}}$ for both models are also shown in each sub-plot.} 
    \label{PCA_PL}
\end{figure*}

We assessed the first ten principal components (PCs) for each sub-band, finding that only the top three were significant. Of these, the first PC dominated, with an explained variance of 58\% to 85\% in Stokes $I$ and 53\% to 77\% for linear polarisation among the top three principal components. Sub-bands A--C exhibited lower explained variances in both Stokes $I$ and linear polarisation, while sub-bands D--H showed higher explained variances, which we hypothesise is due to the influence of residual radio frequency interference (RFI) remaining in the low-frequency data. We show the first PC scores (PC1 scores) across frequency and time in Figure \ref{PCA_PL}. We find that following the event, the PC1 scores show an abrupt increase before gradually decaying. Motivated by this behaviour, we assessed an exponential and a power-law decay model that phenomenologically describe the long-term behaviour of PC1 scores.

The exponential decay model is given by
\begin{equation}
D_e(t) =
\begin{cases} 
    B_\mathrm{pre} & \text{if } t < T_e \\
    A_e \exp\left(-(t - T_{e})/\tau_e\right)+ B_\mathrm{post}, & \text{if } t > T_e 
\end{cases},
\label{eq:expdecay}
\end{equation}
where $B_\mathrm{pre}$ ($B_\mathrm{post}$) is the pre- (post-)event baseline level, $A$ is the exponential amplitude, $T_{e}$ is the event time (MJD 59320; fixed), $\tau$ is the characteristic timescale of the exponential decay. We chose to model the pre- and post-event baseline levels separately to assess the evidence for a permanent reconfiguration of the profile, which would correspond with $B_\mathrm{pre} \neq B_\mathrm{post}$.

The power-law decay model is
\begin{equation}
D_{pl}(t) = 
\begin{cases}
B_\mathrm{pre}, & \text{for } t < T_e \\
A_{pl} \left[1 + \dfrac{t - T_e}{\tau_{pl}} \right]^{-\alpha} + B_\mathrm{post}, & \text{for } t \geq T_e
\end{cases}
\label{eq:PL}
\end{equation}
\noindent
where $A_{pl}$ is the amplitude of the power-law decay at the event time, $\alpha$ is the power-law index, $T_e = 59320$ is the MJD of the profile change event, $\tau_{pl} = 365.25$ is the reference timescale (set to one year), and $B_\mathrm{pre}$ is the baseline offset.

We performed non-linear least-squares optimisation on the exponential decay and power-law models on the PC1 scores using \textsc{Scipy}'s \textsc{curve\_fit} routine, with the Trust Region Reflective algorithm. We set parameter bounds to constrain the fits for both models. In the power law model, we set the bounds of $A$ and $\alpha$ to be positive to avoid non-physical solutions. We also did the same in the exponential model for the $A$ and $\tau$ parameters. We allowed both negative and positive values for both $B_{pre}$ and $B_{post}$. We performed the fits for both Stokes $I$ and linear polarisation. We estimated the uncertainties on the data points using the standard deviation of the pre-event PC1 scores. 

To assess whether the exponential (exp.) or power-law (PL) model better describes the long-term evolution of the profile change, we calculated the reduced chi-squared ($\chi^{2}_{\mathrm{r}}$) statistic to assess overall goodness of fit and the Bayesian Information Criterion (BIC) to enable a direct model comparison between the exp. and PL models. These metrics were calculated for each sub-band and applied to both Stokes $I$ and linear polarisation. 

The BIC values are calculated as:

\begin{equation}
\mathrm{BIC} = k \ln(n) + n \ln\left(\frac{\sum_{i=1}^{n} (P_{i} - D(t_{i}))^2}{n}\right)
\label{BIC}
\end{equation}

\noindent
where $k$ is the number of free parameters in the model, $n$ is the number of epochs, and $P_{i}$ is the principal component score at epoch $t_i$, $D(t_{i})$ is the corresponding model prediction, and the difference of these is the residuals.

We evaluated these metrics using two versions of the PCA timeseries dataset: the full dataset (including both the pre and post-event epochs) and a restricted tail-end subset, beginning at MJD 59800, which was used to assess the asymptotic behaviour of the data while ignoring the complex evolution at early times post-event. To determine the preferred model, we evaluated the difference in BICs:
\begin{equation}
\Delta \mathrm{BIC} = \mathrm{BIC}_{\rm Exp} - \mathrm{BIC}_{\rm PL},
\label{DeltaBIC}
\end{equation}
where the subscripts denote the BIC scores of the exp. and PL models, respectively. A negative value of $\Delta \mathrm{BIC}$ indicates that the exponential model is preferred over the power-law model, while a positive value is indicative of the power-law model being favoured.

\begin{figure*}
    \centering
    \includegraphics[width=\linewidth]{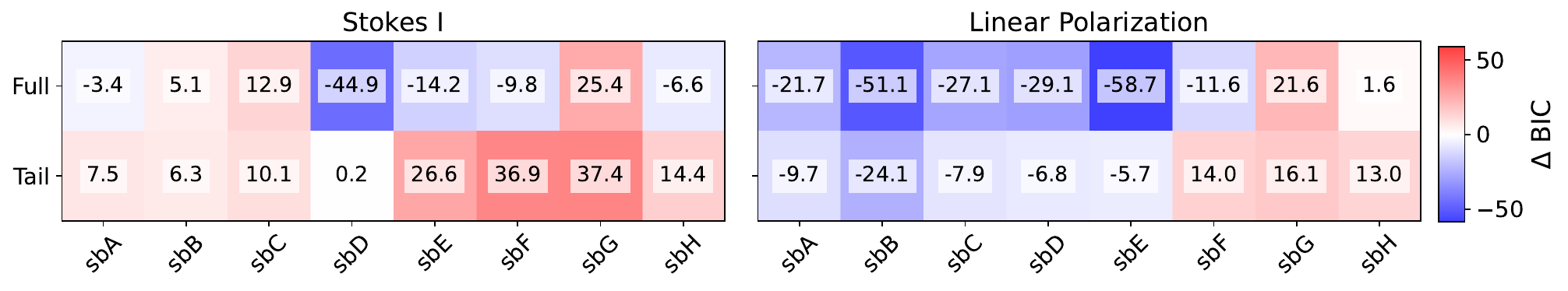}
    \caption{Difference in BIC scores indicating model preference for either the exponential (Exp) or power law (PL) fit to the PC scores for the first principle component of the Stokes $I$ and linear polarisation residuals. A negative $\Delta\mathrm{BIC}$ value indicates that the exponential model is favoured, whereas a positive value indicates the power-law model is favoured. The full dataset encompasses both pre and post-event epochs. The tail dataset is a restricted subset that commences at MJD 59800 where the fitted curve appears to flatten. In the Stokes $I$tail dataset, the power-law model is dominant, whereas support for either model is mixed for linear polarisation. Here, $\Delta\mathrm{BIC}=\mathrm{BIC}_{\mathrm{Exp}}- \mathrm{BIC}_{\mathrm{PL}}$. For clarity, we indicate $\Delta \mathrm{BIC}$ for each data sub-set on each cell.}
    \label{PCA_models}
\end{figure*}

While neither model perfectly captures the full evolution of the profile change (as seen from the $\chi^{2}_{\mathrm{r}}$ values in Figure \ref{PCA_PL}), the PL model is predominantly supported in Stokes I, particularly in the tail-end dataset. We summarise the $\Delta \mathrm{BIC}$ across sub-bands, polarisation, and data sub-samples in Figure \ref{PCA_models}.

In Figure \ref{PCA_PL}, along with the PC1 scores themselves, we show the best-fit power-law models as a function of time, across frequency and polarisation. In the lower-frequency sub-bands, the scores and fitted curves are very similar for Stokes $I$ and linear polarisation. However, above $\sim1300$\,MHz, the separation between PC1 scores for total intensity and linear polarisation grows, with Stokes $I$exhibiting higher PC scores than linear polarisation. Notwithstanding this separation, the post-event PC scores grow in magnitude as a function of frequency both for Stokes $I$ and linear polarisation.

\begin{figure}
    \centering
    \includegraphics[width=\columnwidth]{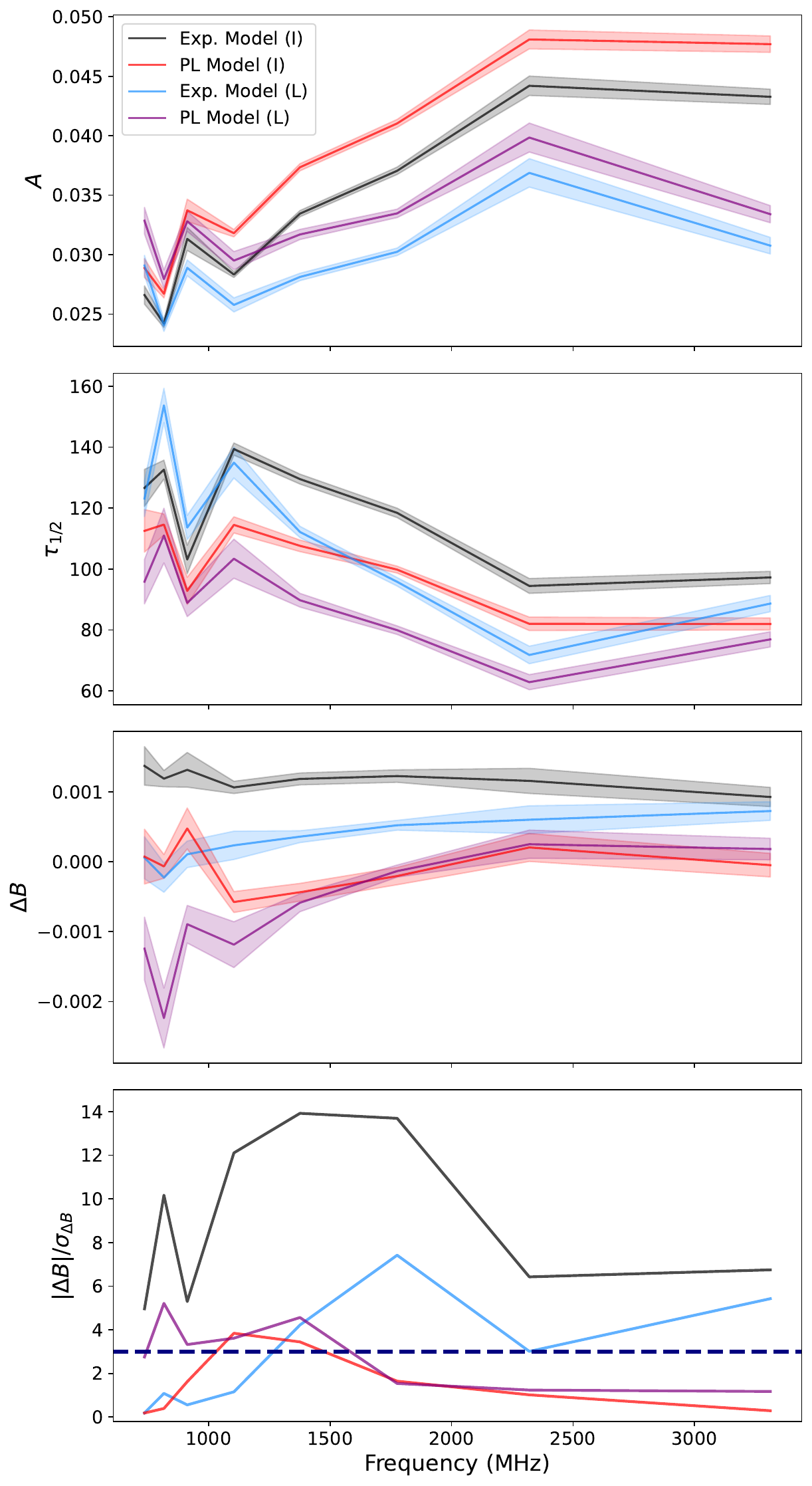}
    \caption{
    Evolution of fitted parameters to Stokes $I$ and linear polarisation PC scores as a function of frequency for the exponential-decay (Exp) and power-law (PL) models (see equations \ref{eq:expdecay} and \ref{eq:PL}). First Panel: model amplitude $A$; Second Panel: half-life decay times $\tau_{1/2}$; Third Panel: $\Delta B$ -- the difference between the post-event and pre-event baselines; Fourth Panel: S/N of $\Delta B$, with the dashed blue line indicating the $3\sigma$ significance threshold. Shaded regions around each line represent the uncertainties on the fit parameters. Black (light blue) indicate the model fits for the Exp. model in Stokes $I$ (linear polarisation), while red (purple) indicate the model fits for the PL model in Stokes $I$ (linear polarisation).}
    \label{PCA_params}
\end{figure}

This behaviour is reflected in the top two panels of Figure \ref{PCA_params}, which shows the amplitude and derived half-life from both the exponential decay and power-law model fits to the PC scores as a function of frequency. The model amplitudes show a gradual increase with frequency, though the increase is non-monotonic. Similarly, the derived half-lives decline gradually (but non-monotonically) with frequency. The parameters follow similar trends across the different polarisations and models, although with offsets between them. 

Both our exp. and PL models include a free parameter $B_\mathrm{post}$ to account for a non-zero asymptotic offset from the pre-event baseline value $B_\mathrm{pre}$. We measure the absolute value of this offset as $\Delta B = |B_{\mathrm{pre}} - B_{\mathrm{post}}|$, and assess the significance of this offset by quantifying the offset S/N as $|\Delta B|/\sigma_{\Delta B}$ where $\sigma_{\Delta B}$ is the uncertainty of the offset $\Delta B$. We adopt a threshold of 3 as being indicative of a permanent reconfiguration of the profile. We show the behaviour of $\Delta B$ in the lower two panels of Figure \ref{PCA_params}. We found that a significant non-zero offset is consistently preferred under the exponential model. However, as discussed above, the power-law model is predominantly supported in Stokes I. In the case of the power-law fits, only some sub-bands showed significant evidence for a non-zero offset: in Stokes I, only sub-band D and sub-band E showed offsets  exceeding 3$\sigma$ significance ($3.8\sigma$ and $3.4\sigma$ respectively), whereas in linear polarisation, sub-bands B -- E showed offsets with significances ranging from $3.2$ -- $5.2\sigma$. In light of these mixed results, it is unclear whether the profile, overall, is fully recovering to its pre-event state, or whether some components may not fully recover. Ongoing long-term monitoring will be required to answer this question more comprehensively.

\subsection{Polarimetric analysis of the profile change}
\label{sec:OPM_changes}

Pulsar emission is intrinsically polarised and directional, which results in complex, frequency-dependent polarisation profiles \citep{1998ApJ...501..286X,2015MNRAS.449.3223D,2021MNRAS.505.4483S,2023MNRAS.520.4961O}. A key feature of the polarisation in some pulsars is the presence of OPMs, which are thought to be generated as a result of two different modes of wave propagation in the pulsar magnetosphere: the ordinary (O) mode -- in which the wave propagation runs parallel to field lines, and the extraordinary (X) mode -- in which wave propagation is orthogonally aligned to the orientation of field lines. The intrinsic emission contains components of both these modes. As the waves travel through the birefringent plasma, the O and X mode propagate with different velocities, resulting in a relative phase delay. Upon leaving the plasma, these components recombine and the observed dominant mode depends on the magnitude of this phase delay. As a result, the observed PA can exhibit sudden 90-degree jumps, as one mode becomes dominant over the other across the pulse longitudinal phase. For a comprehensive summary on OPMs, see   \cite{1997A&A...327..155G,1998A&A...336..209V,2001A&A...378..883P,2009MNRAS.392L..60K,2015A&A...576A..62N,2016A&A...593A..83D,2014MNRAS.441.1943W,2017MNRAS.472.4598D,2023MNRAS.525..840O}.

Like the total-intensity profiles, individual pulses exhibit variability, but the integrated polarisation and PA profiles in MSPs are very stable -- with the boundaries of the two modes clearly distinguished. The OPMs also show frequency dependence in their location and extent across the profiles \citep{2006A&A...448.1139S}. The superposition and coherent/incoherent mixing of both the O and X mode emission is thought to lead to depolarisation \citep{1984ApJS...55..247S}, while the partially coherent mixing of modes has been proposed as an origin of pulsar circular polarisation \citep{2023MNRAS.525..840O}. Fractional polarisation can therefore provide insights into the overall polarisation profile configuration and the dominance of either of the OPMs across the pulse phase: a high fractional linear polarisation indicates the dominance of one of the modes, whereas low or zero fractional polarisation may indicate mode mixing.

\subsubsection{Variability in OPM intensities}
\label{sec:OPM_changes_opmflux}
\begin{figure*}
    \centering
    \includegraphics[width=\linewidth]{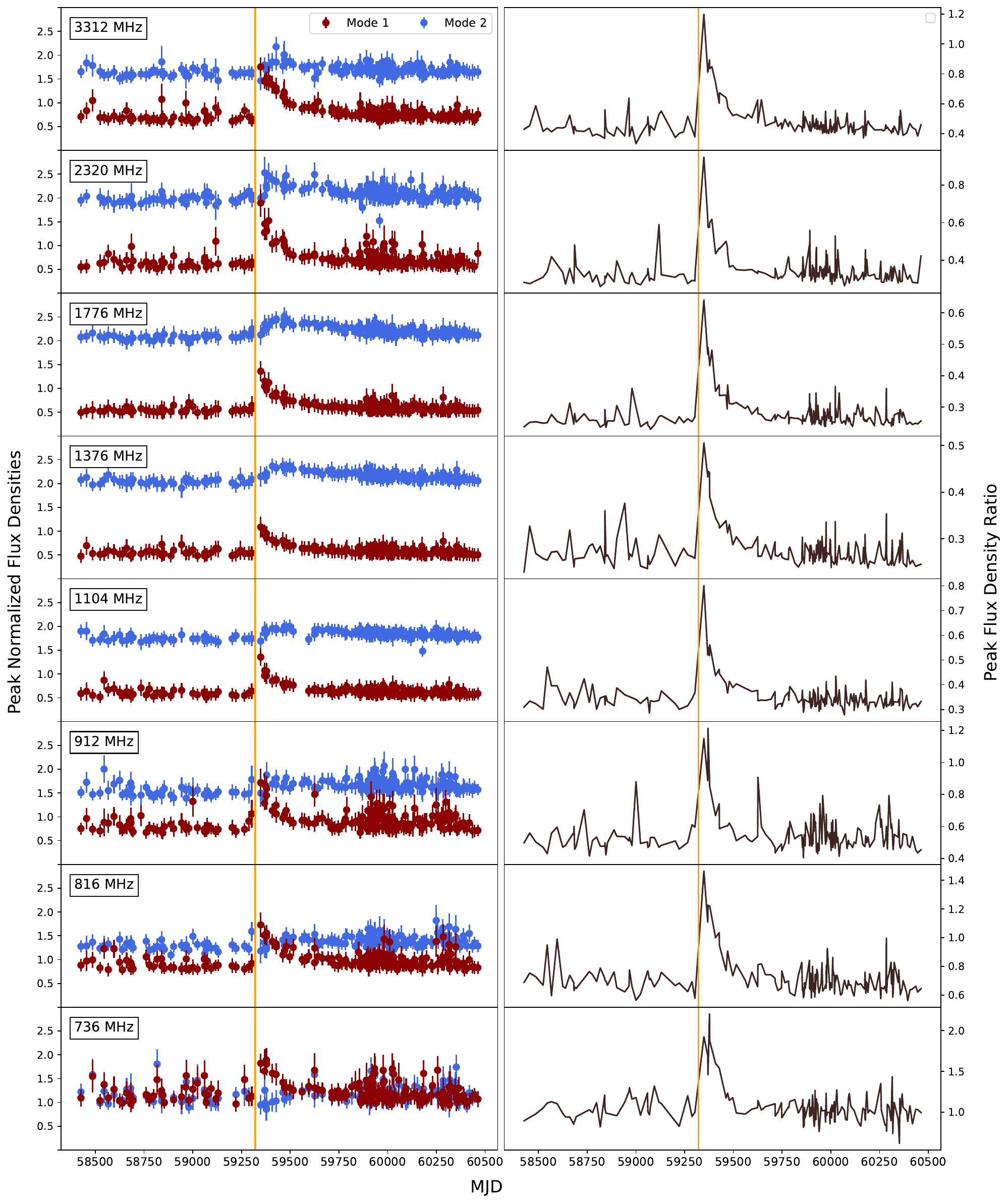}
    \caption{Left Column: normalised flux densities of the two profile peaks located in the central phase bins of the linear polarisation profile that represent the two OPM modes, plotted as a function of time and in different frequency bands. The red points indicates the first mode, while blue points indicate the second mode. Right: Flux density ratio of the same two central OPM components. The vertical orange line represents the profile change date (MJD 59320/59321).}
    \label{OPM_FluxAmp}
\end{figure*}

J1713 exhibits several OPM transitions, which demarcate four linear polarisation sub-components that we annotate in Figure \ref{IPALV_3epoch}. To analyse how the profile change event affected these components, we first measure the peak normalised flux density of the profile components corresponding to these two OPMs over time across all UWL sub-bands. These are presented in the left column of Figure \ref{OPM_FluxAmp}. Specific patterns emerge in the behaviour of the OPM sub-components following the profile change event. The first OPM sub-component (corresponding to negative PA values in Figure \ref{IPALV_3epoch}), for example, shows a significant increase in flux density immediately after the event across all sub-bands, before undergoing an exponential-like decay. The second OPM sub-component (corresponding to positive PA values) is more stable, but exhibits a small but visible amplitude increase immediately after the event. We also measured the ratio of the peak flux densities of the two OPM components over time, for each frequency sub-band. As illustrated in the right column of Figure \ref{OPM_FluxAmp}, the profile change event produced a pronounced spike in the flux density ratio, followed by a gradual decay.

\subsubsection{Variability in Linear polarisation PAs}
\label{subsubsec:PA_var}
\label{sec:OPM_changes_PA}
\begin{figure*}
    \centering
    \includegraphics[width=\linewidth]{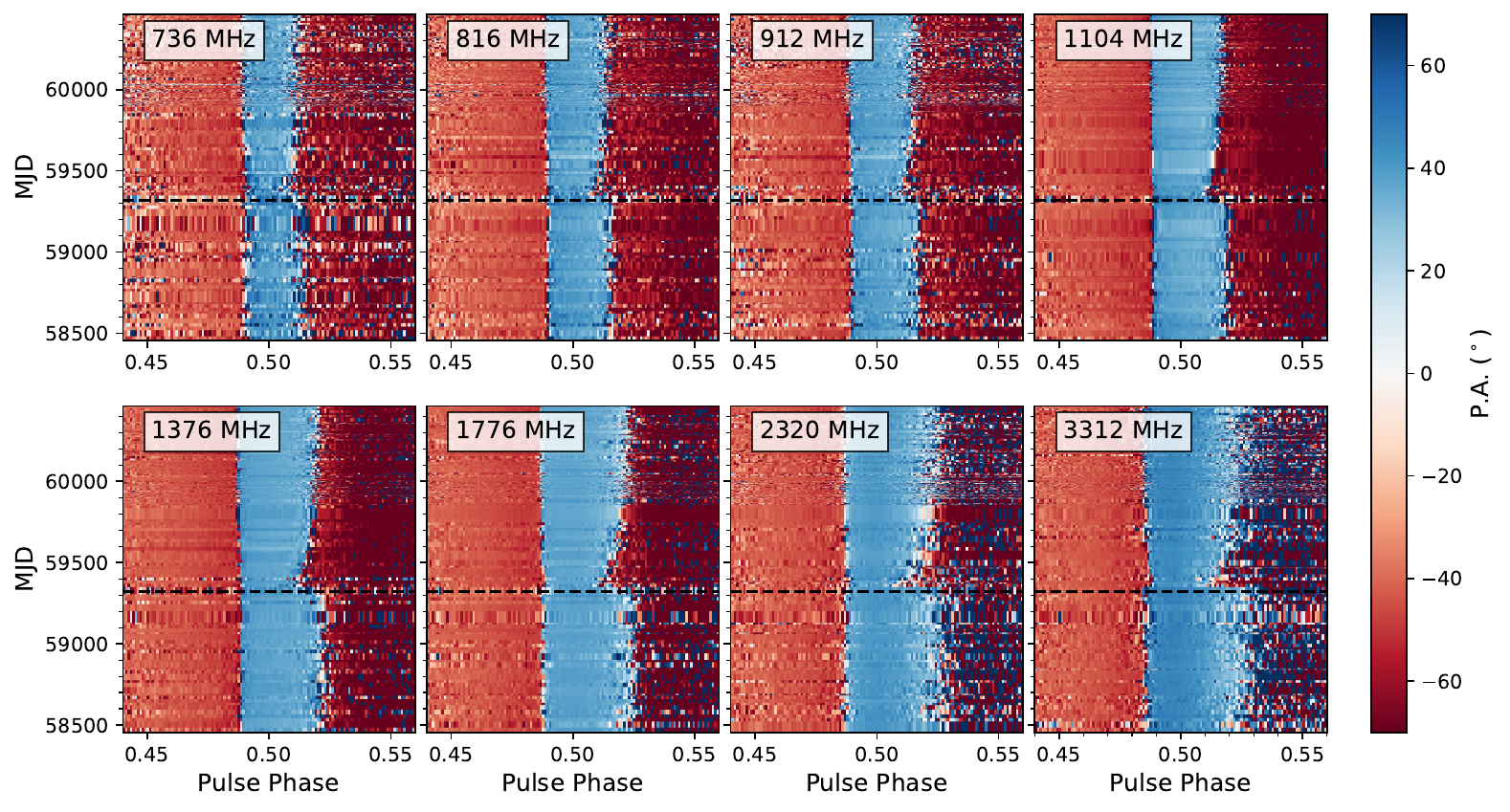}
    \caption{Time-series of the PA of the central profile sub-components (OPM 1.1 (light red), OPM 2.1 (light blue), OPM 1.2 (dark red)) over the 8 frequency sub-bands, which are labelled in the top left of each sub-figure. We have flagged out PAs in phase bins where the Stokes $I$signal-to-noise was lower than 2. The black dashed line indicates the profile change event date. The colourmap spans from $-70^\circ$ (dark red) to $70^\circ$ (dark blue).}
    \label{PA_waterfall}
\end{figure*}

We examined PA variability using the ephemeris-aligned profile dataset. Figure~\ref{PA_waterfall} presents this evolution over time, where we present the PA across OPM components 1.1, 2.1, 1.2 for each sub-band. Notably, the first sub-component (1.1) remains remarkably stable across all epochs, and in particular, its right-hand boundary, where the mode transition to sub-component 2.1 occurs does not appear to shift significantly in pulse phase, nor do the PA values themselves vary. The fact that the PAs do not appear to vary in this leading edge of the profile, while the (linearly polarised) flux density dramatically varies, highlights that this leading edge of the profile is dominated by one orthogonal mode of emission. However, the PA boundary between 2.1 and 1.2 shows significant variation following the profile change event. Specifically, PAs on the left boundary of 1.2 extend leftward, reflecting a contracted phase width for sub-component 2.1. The magnitude of this leftward incursion of component 1.2 into 2.1 increases with frequency, demonstrating a frequency-dependent polarimetric effect of the profile change event. 

\begin{figure*}
    \centering
    \includegraphics[width=\linewidth]{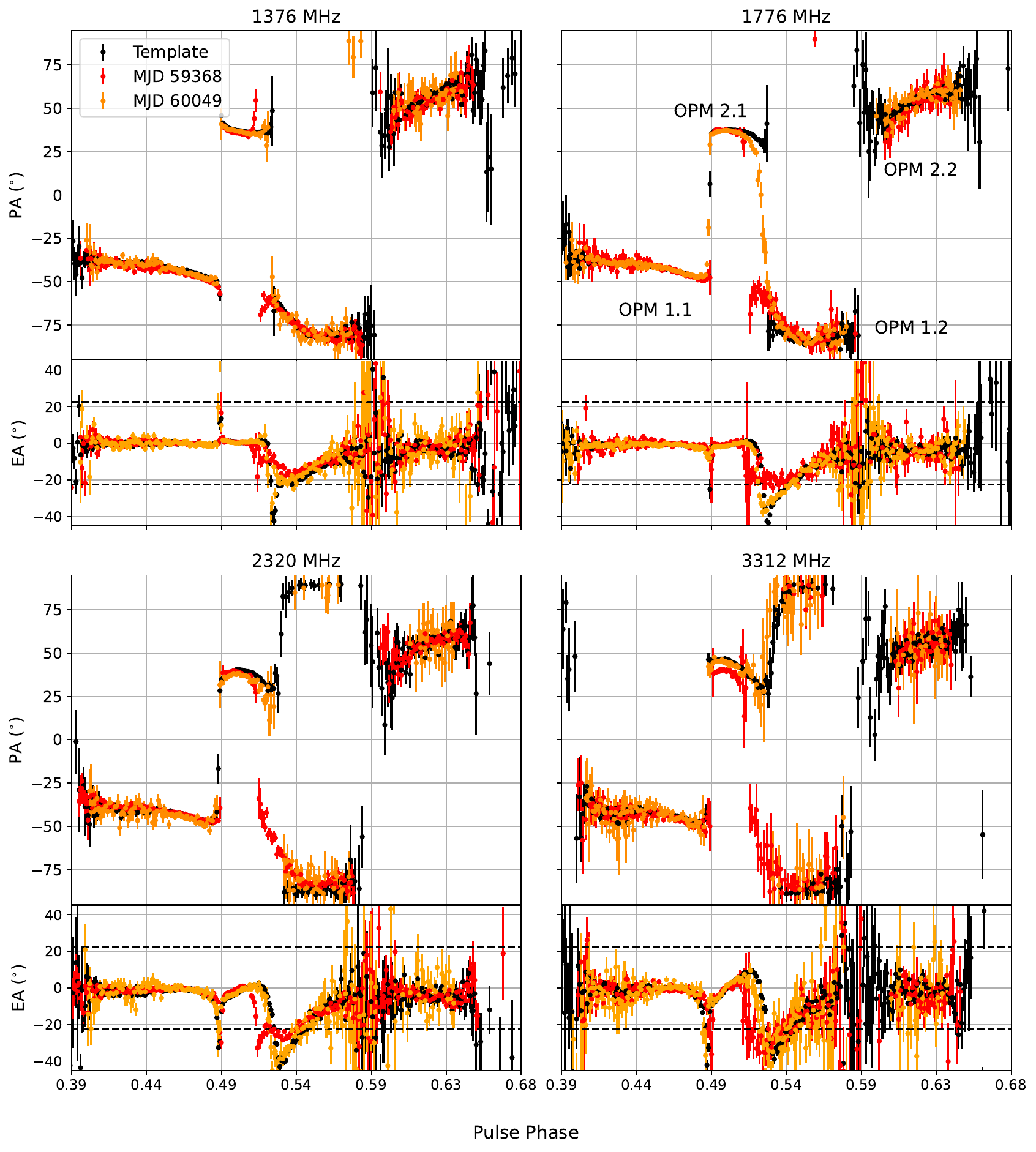}
    \caption{Top row: PA as a function of pulse phase in the four highest sub-bands, showing template data in black, an early post-event epoch ($\sim$50 days after the event; MJD 59368) in red, and a late epoch ($\sim$two years post-event; MJD 60049) in orange. A distinct modal bridge emerges between sub-components 2.1 and 1.2, suggesting a complex interaction between the OPMs. Sub-component 1.1 remains stable throughout the profile change event. Bottom row: Ellipticity Angle (EA) as a function of pulse phase for the same four sub-bands, showing the template in black, the same early post-event epoch (MJD 59368) in red, and the later post-event epoch in (MJD 60049) in orange.}
    \label{Modal_Bridges}
\end{figure*}

The impact of this shift in OPMs is also evident in Figure \ref{Modal_Bridges}, in both the Stokes $I$ and linear polarisation profiles immediately after the event. In Stokes $I$, profile component C narrows significantly, consistent with the contraction of OPM component 2.1. Profile component D, associated with OPM component 1.2, shows frequency-dependent variability, with the formation of a flat, `L-shaped' plateau-like structure in the lower sub-bands and more gradual gradients in the higher sub-bands. In the linear polarisation profile, the central component narrows only slightly, but the higher sub-bands corresponding to OPM component 1.2 increase in flux density, matching the amplitude seen in lower sub-bands prior to the event, no longer exhibiting the clear separation of frequency-dependent amplitudes for this component. 

Figure \ref{Modal_Bridges} illustrates the PA sub-components for the four highest sub-bands, comparing the template (black) with the two post-event reference epochs shown in Figure \ref{IPALV_3epoch}: MJD 59368 (red; 210603) and MJD 60049 (orange; 230415), along with the ellipticity angle (EA), defined as $1/2 \arctan 2(V/L)$. Examining each sub-component individually confirms that the first component (1.1) remains stable across both time and frequency. In contrast, the third component (1.2) exhibits a pronounced deviation from the template at both epochs, forming a ``modal bridge'' (i.e. a phase resolved OPM transition) in which the PA values extend leftward and upward toward the second component (2.1), which suggests a complex interaction between the two modes of emission. At similar pulse phases, the EA decreases below $-22.5^\circ$, indicating that $|V| > L$ in these regions. The second component itself displays a clear contraction, immediately post-event. These results represent the first evidence of a modal bridge forming after a profile change event in any MSP. This modal restructuring occurs in pulse phase bins where the profile shape itself is altered, indicating a direct link between OPM phenomenology and the overall profile changes. Additionally, these features persist well into the later post-event epoch (MJD 60049), highlighting the long-term nature of this event.

We also note that Figure \ref{IPALV_3epoch} suggests an apparent shift of the modal bridge towards lower frequencies at later epochs. However, analysis across a wide range of post-event epochs shows that this behaviour is not consistent, with several later epochs exhibiting modal bridge structures in both high and low frequency bands. Our ability to probe this behaviour in more detail is limited by low S/N in many epochs, and we therefore suggest detailed investigation of this effect should be carried out with more sensitive telescopes like MeerKAT or FAST \citep{2025MNRAS.536.1467M,2025A&A...695A.173X}.

\subsubsection{Variability in fractional polarisation}

\begin{figure*}
    \centering
    \includegraphics[width=16cm]{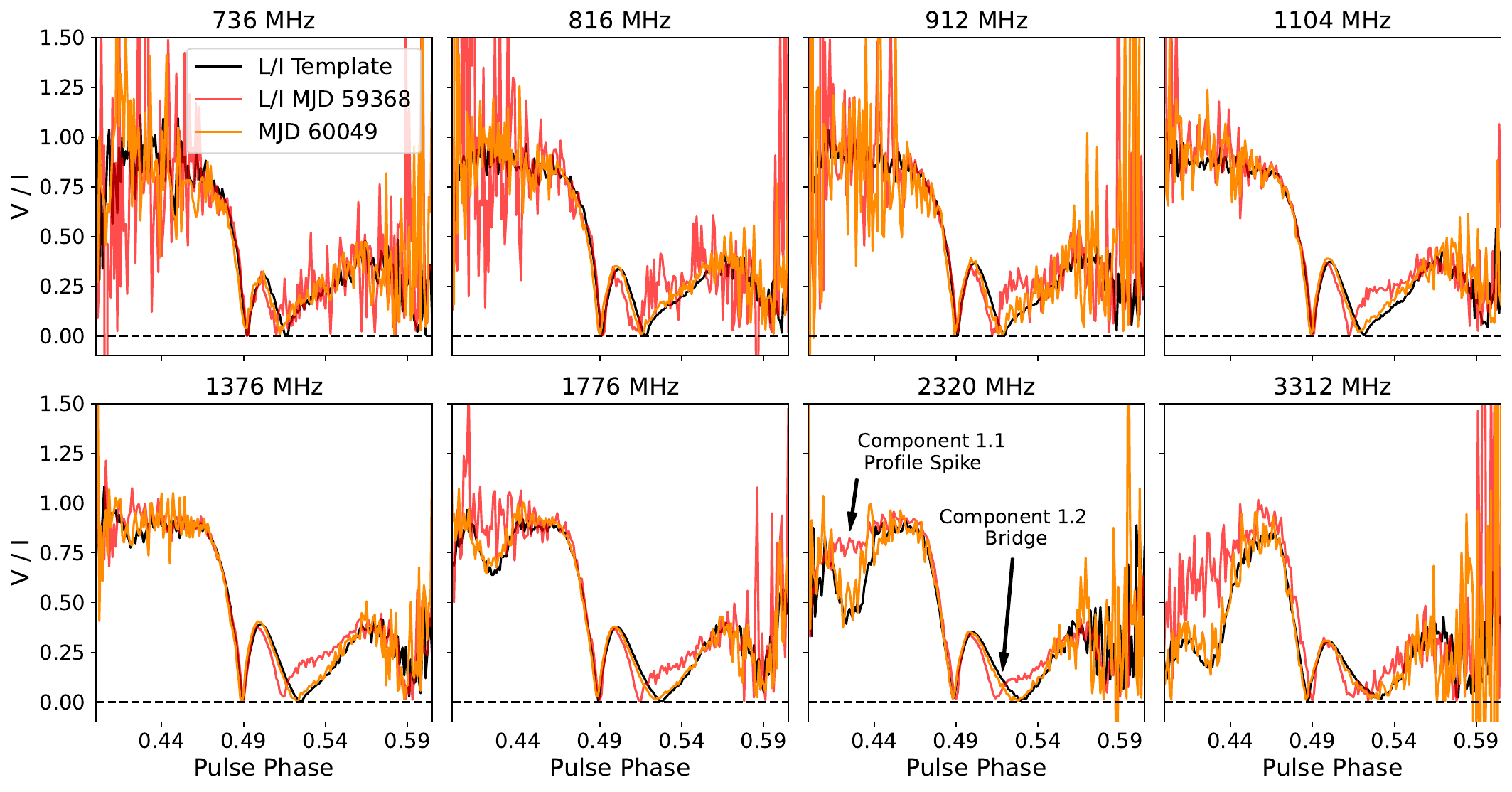}
    \caption{Fractional linear polarisation over central bins of pulse phase as a function of both frequency and time. The black curve represents the pre-event template, the red curve represents our first reference epoch (MJD 59368) and the orange curve represents our second reference epoch (MJD 60049). In the bottom left sub-plot, we annotated the sudden increase in $L/I$ in this region of the profile, as well as the location of the modal bridge.}
    \label{LinFrac_phase}
\end{figure*}
\label{subsubsec:fracpol_var}

We also investigated the behaviour of fractional polarisation over time, examining phase-averaged OPM profile components over frequency, and the profile overall in a fully phase-resolved fashion. Figure~\ref{LinFrac_phase} presents the fractional linear polarisation as a function of pulse phase across each sub-band for the template (black), and two high S/N post-event reference epochs: MJD 59368 (red) and MJD 60049 (orange).

The template shows a central peak in $L/I$ corresponding to the main pulse component (profile component 'C' in Figure \ref{IPALV_3epoch}), with steep declines to near-zero flux densities on either side (pulse phase 0.49--0.52) corresponding to the transitions between OPM sub-components. In the early post-event epoch (MJD 59368; red curve in Figure \ref{LinFrac_phase}), the first depolarisation boundary between OPM sub-components 1.1 and 2.1 remains relatively stable across frequency, with a slight deviation at the highest sub-band (sbH; 3312\,MHz). In contrast, the second depolarisation boundary between sub-components 2.1 and 1.2 shifts leftward in phase following the profile event, with the shift becoming more pronounced at higher frequencies. This frequency-dependent shift aligns with the modal bridge observed in the PAs, particularly around 2320\,MHz (sbG), where we annotate this feature as the component 1.2 bridge (see Figure \ref{Modal_Bridges}). At these phases, the EA also decreases below $-22.5^\circ$, indicating that circular polarisation dominates the polarised intensity. By the later reference epoch (MJD 60049; orange curve in Figure \ref{LinFrac_phase}), the fractional linear polarisation closely resembles the template, indicating a gradual recovery of the polarisation over time. 

We also note the emergence of an additional fractional polarisation feature in the highest three sub-bands around pulse phase $\sim$0.40 -- 0.44 (Figure \ref{LinFrac_phase}), where the MJD~59368 profile (red curve) exhibits a significant increase in fractional linear polarisation relative to the template. We annotate this as the component 1.1 profile spike, corresponding to the emergence of a peak in linearly polarised intensity within phases 0.40 -- 0.44 in the highest frequency sub-bands, as seen in Figures~\ref{IPALV_3epoch}, \ref{fig:waterfalls_Eph} and \ref{fig:waterfalls_FFT}. We note that this polarisation feature corresponds with the emergence of a persistent new, faint profile feature most apparent in the higher-frequency sub-bands of both Figure~\ref{fig:waterfalls_Eph} and Figure~\ref{fig:waterfalls_FFT}, around the location of profile component `A' labelled in Figure \ref{IPALV_3epoch}. Despite the significant change in fractional polarisation in this phase range, the corresponding linear polarisation PA remains unaffected.

\begin{figure*}
    \centering
    \includegraphics[width=\linewidth]{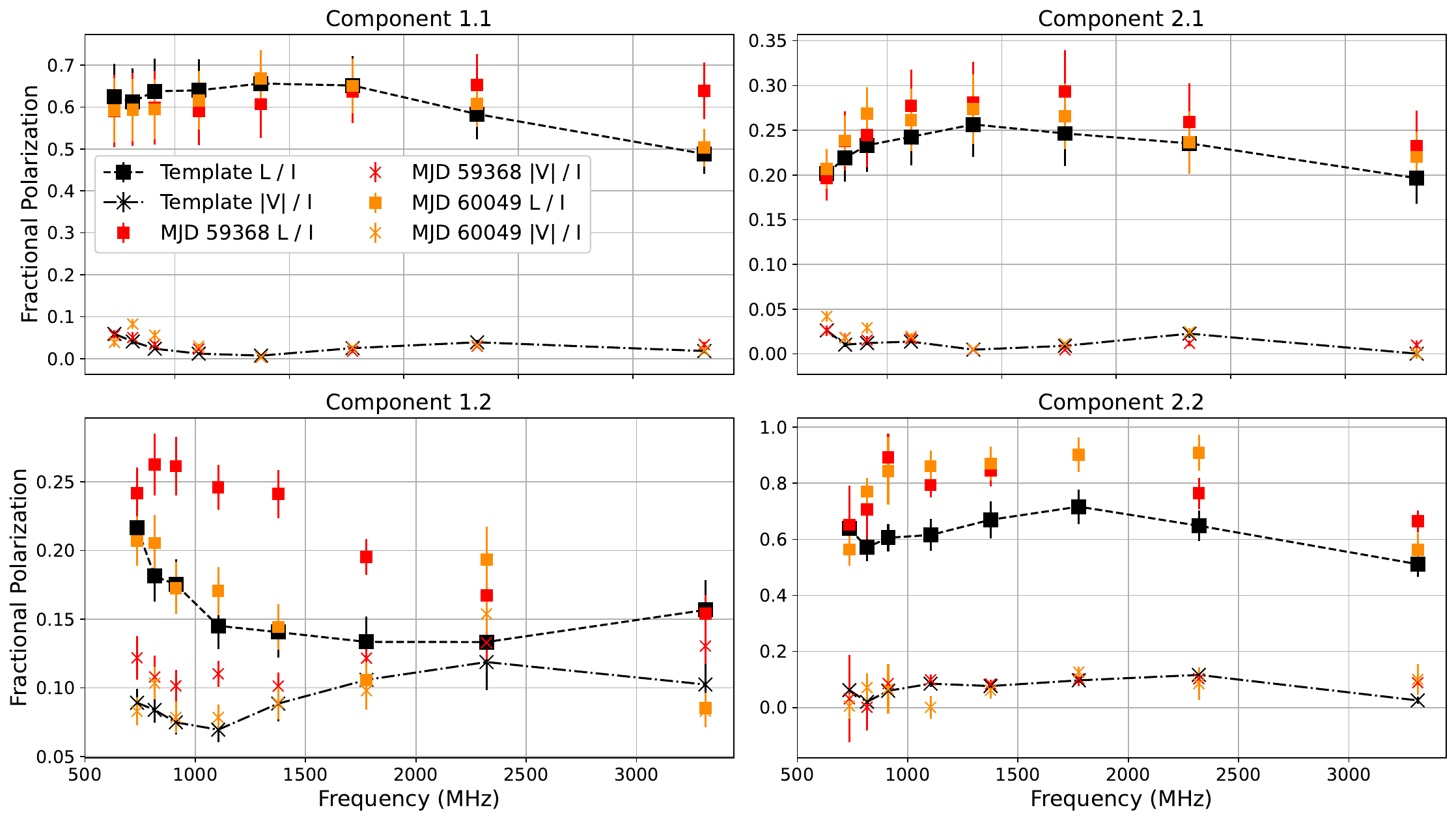}
    \caption{Fractional linear (squares) and fractional circular polarisation (crosses) per OPM component as a function of both frequency and time. The black data points represents the pre-event template, the red data points represents our first reference epoch (MJD 59368) and the orange data points represents our second reference epoch (MJD 60049).}
    \label{OPM_FracPols}
\end{figure*}

We show the average fractional polarisation as a function of frequency, for both linear and circular polarisation (Stokes V), across different OPM sub-components in Figure~\ref{OPM_FracPols}. We compute the average quantities within the respective pulse phase boundaries of each sub-component, which we defined by the continuous extent of the PA clusters seen in e.g. Figure \ref{IPALV_3epoch}. The results are shown for all sub-bands and three key epochs: the pre-event template (black), and our two reference epochs as highlighted previously in Figure \ref{Modal_Bridges} and \ref{LinFrac_phase}, 50 days (red) and $\sim$2 years (orange) post event. Fractional linear polarisation is indicated with square markers and circular polarisation with cross markers. Components 1.1 and 2.1 (representing the leading edge of the profile) show only mild variability across these examples. In contrast, components 1.2 and 2.2 (on the profile trailing edge) exhibit significant variability in the spectral behaviour of fractional polarisation. For component 1.2, the early post-event epoch shows a non-monotonic frequency dependence in fractional linear polarisation, with most sub-bands eventually recovering to near-template values by the late epoch, though some tension remains in the highest frequency bands. Stokes $V$ also displays notable variation in this component. In component 2.2, while Stokes $V$ remains relatively unaffected, fractional linear polarisation significantly varies immediately after the event and had not yet recovered by MJD 60049.

As shown in Figure~\ref{OPM_FracPols}, we also observe the fractional polarisation of the leading sub-components -- corresponding to the first half of the polarisation profile -- remains relatively stable during and well after the event. In contrast, the trailing sub-components exhibit significant deviations from the pre-event template, even persisting in the later reference epoch. A similar pattern is seen in Stokes V, which is minimally affected in the leading half of the profile but displays greater variability in the trailing region. 

\section{Discussion}
\label{sec:discussion}

\subsection{The origin of the profile change event}
Discrete profile changes are rare amongst the MSP population. A comparable case to the J1713 event considered in this paper is the profile change event in PSR~J1643$-$1224, where a new profile component appeared in observations taken at $\sim 3000$\,MHz and $\sim 800$\,MHz \citep{2016ApJ...828L...1S,2018ApJ...868..122B}, and decayed over approximately four months. However, one key distinction from the PSR~J1643$-$1224 case is that its new component was unpolarised, whereas the variations we report in the J1713 profile change event exhibit clearly distinguished frequency-dependent variations to all Stokes parameters.

Long-lasting profile variations as a result of a discrete profile change were also reported in PSR~J0437$-$4715 \citep{2021MNRAS.502..478G}, with profile residuals persisting for over a year. Similarly, \cite{2018ApJ...868..122B} also documented profile variability in PSR~J2145$-$0750 and PSR~B1937$+$21, although these were attributed in part to instrumental systematics, such as polarisation miscalibration, RFI, as well as signal propagation effects such as scintillation and scatter broadening. Outside of these examples, profile change events in the MSP population that exhibit such pronounced variations in the profile shape structure and multi-year recovery times are non-existent.

While the physical mechanism that triggered the April 2021 profile change remains unknown, the differential response between the two OPMs is suggestive of environmental changes in their magnetospheric origin, either close to the emission region \citep{1979ApJ...229..348C} or at higher altitudes \citep{2001A&A...378..883P}.  This is also supported by the frequency-dependence of the fitted parameters to the PCA scores seen in Figure \ref{PCA_params}, which show that the amplitude of the profile change increases with frequency, while the recovery timescale declines with frequency. This is inconsistent with line-of-sight propagation effects, which therefore makes an intrinsic magnetospheric origin more favourable. However, pinpointing the exact in-situ cause of the event remains challenging, owing to the complex and poorly understood nature of MSP magnetospheres \citep{1998ApJ...501..286X,1999ApJS..123..627S,2025A&A...695A.173X}. 

One possibility is that the event was caused by a sudden reconfiguration of the MSP magnetic field, as suggested previously \citep{2021MNRAS.502..478G,2021MNRAS.507L..57S,2024ApJ...964..179J,2024ApJ...961...48W}. Such a reconfiguration would alter the emission geometry and as such, the pulse profile. It may also alter the relative strengths of the OPMs in different pulse phase regions, therefore shifting the pulse phases where depolarisation occurs as each mode transitions in dominance as seen in Figure \ref{PA_waterfall} and \ref{LinFrac_phase}. The observed profile shape change would then exhibit temporal evolution as the field returned to its prior configuration over time. This picture is consistent with our observation of broad-band variability in the profile and its OPM components, which we note has not been previously reported in connection to this event or other MSP profile changes. Whilst not directly associated with discrete profile changes, \cite{2010MNRAS.408L..41T} and \cite{2017MNRAS.469.2049Y} highlight examples of the variations to pulsar emissions resulting from changes in the magnetospheric configuration and emission geometry.
 
Additionally, reconfigurations of magnetic field lines are intrinsically linked to a redistribution of plasma density, since active field lines above polar caps -- where pair production occurs and emissions are generated -- are determined by the global magnetic field structure \citep{2008ApJ...683L..41B,2010MNRAS.408.2092T}. As such, a redistribution of plasma density could modify the refractive indices for the O and X modes, altering the degree of mode separation, augmenting the mode mixing and leading to observable changes in the relative dominance of either OPM \citep{2015MNRAS.448..771W,2015A&A...576A..62N}. This scenario would be consistent with the non-monotonic frequency-dependence observed in the fractional polarisation for certain regions of the profile (marked by the boundaries of the OPM sub-components, see Figure \ref{OPM_FracPols} in particular sub-component 1.2), as birefringence effects scale with observing frequency \citep{1997ApJ...475..763M}. Further, the enhanced ellipticity angle at the pulse phases corresponding to new ``modal bridges'' detailed in Section \ref{subsubsec:PA_var} is suggestive of partially-coherent mode combination \citep{2023MNRAS.525..840O}, consistent with changes in the properties of the magnetospheric plasma.

One postulated driver of changes to the magnetospheric plasma is the evaporation of an asteroid as it enters the pulsar magnetosphere, which may cause profile shape changes and alter the pulsar spin properties \citep[e.g.][]{2008ApJ...682.1152C,2014ApJ...780L..31B}. 
We have not yet carried out a detailed investigated whether the April 2021 profile change was accompanied by variations in the MSP’s rotational period or period derivative before and after the event. This analysis will be presented in an upcoming paper focused on the timing impacts of the event. To date, there have been no reports of changes in the spin properties of J1713 following the event, nor is there any evidence of a large glitch associated with this event.

As discussed in Section \ref{sec:results} and illustrated in Figures~\ref{fig:waterfalls_Eph} and \ref{fig:waterfalls_FFT}, the profile residuals persist through to the latest epoch in our dataset, irrespective of the alignment method used. The absence of similar residuals in the pre-event data under both alignment approaches suggests that this is not a systematic artifact, but a genuine consequence of the April 2021 profile change. 
This raises a key question: has the pulse profile of J1713 undergone a permanent reconfiguration, or is it still gradually returning to its original configuration? If the former, this would mark only the second such case among MSPs -- after the profile reconfiguration in PSR~J1643$-$1224 reported by \citet{2016ApJ...828L...1S}. However, as outlined in \ref{PolVar} the results of model fitting to the decaying residuals are not conclusive as to whether the profile has permanently reconfigured or is still recovering. Continued monitoring and analysis will be required to resolve this.

Overall, our findings strongly disfavour external or interstellar propagation-related origins for the April 2021 profile change. We find no evidence for intrinsic emission phenomena such as nulling, mode-changing, or glitching, and find no evidence of signal propagation effects including scattering, pulse broadening, or plasma lensing. Pulse-to-pulse variability (jitter) and scintillation are well-characterised for J1713 \citep{2012ApJ...761...64S,2014ApJ...794...21D,2002ApJ...581..495B, 2018ApJ...868..122B} and do not account for the long-duration, structured profile changes observed in this event.

\subsection{Implications for pulsar timing arrays and future work}
J1713 is a high-priority MSP for all global PTA collaborations. However, the April 2021 profile change event introduced large, frequency-dependent timing offsets \citep{2024ApJ...972...49L} that led to the exclusion of post-event data from PTA datasets \citep{2023A&A...678A..50E,2023RAA....23g5024X,2023ApJ...951L...6R,2023ApJ...951L...8A,2025MNRAS.536.1467M,2025arXiv250616769R}. The exceptional long-term timing stability of J1713 prior to the event means that development of methods to correct for the profile change and enable continued use of this MSP in these high-precision timing campaigns, such as using stable features of the polarisation, should be a priority. While the J1713 event studied in this work is the largest example of its kind for MSPs, these improved techniques to correct for profile changes will prove valuable in the era of highly sensitive observations with FAST and the SKA, which will be capable of detecting smaller profile changes compared to the current catalogue of MSP profile changes.

Rapid-response follow-up observations to profile change events would have scientific benefits beyond potential corrections to timing -- in particular, for understanding the phenomenology of pulse shape changes at early times, and subseqeunt insights e.g. into the magnetosphere dynamics driving them \citep{2011MNRAS.411.1917W, 2017MNRAS.469.2049Y,2024MNRAS.528.3833F}. Additionally, some MSPs exhibit thermal X-ray hotspots near their magnetic poles, caused by returning currents heating the neutron star surface \citep{2006ApJ...638..951Z,2019A&A...627A.141W}. If profile changes are indeed driven by magnetospheric field-line reconfigurations, corresponding variability in these X-ray hotspots may also be detectable, which may present an opportunity for X-ray instruments such as XMM-Newton or NICER \citep{2019ApJ...887L..24M,2019ApJ...887L..21R,2021ApJ...918L..26W}.

Another important avenue for future investigation is the effect of profile change events on single-pulse behaviour, with a particular focus on their polarimetry. For example, during a serendipitously captured glitch in the Vela pulsar, \citet{2018Natur.556..219P} observed a dramatic quenching of the the Stokes $I$ flux density at the moment of the event, accompanied by quenched linear polarisation in the pulses immediately post-glitch, along with variability in the pulse shape both before and after the glitch. This highlights the potential of single-pulse observations as a diagnostic tool during sudden state changes.

\section{CONCLUSION}
\label{sec:conclusion}
We conducted a comprehensive polarimetric analysis of the PSR~J1713$+$0747 April 2021 profile change event spanning three years of observations with the UWL receiver on Murriyang, providing continuous bandwidth from 704\,MHz to 4032\,MHz. Our study was motivated by the insights that such a wide bandwidth, along with high polarimetric fidelity could provide.

Consistent with previous studies, we conclude that the PSR~J1713$+$0747 April 2021 event resulted from a disturbance in the MSP’s magnetosphere, rather than propagation effects through the interstellar medium. In our case, this is supported by the broad bandwidth and long timescale (half-life $\sim$160 days) of the profile change. This is further supported with the distinct polarimetric behaviour, such as a stable PA OPM transition in the profile leading-edge seen in Figure \ref{PA_waterfall}, whereas the trailing edge of the profile displays shifts in the location of OPM transitions along with greater variability in the fractional polarisation (Figure \ref{LinFrac_phase} and \ref{OPM_FracPols}). Indeed, the relative intensity of the orthogonal modes (Figure \ref{OPM_FluxAmp}) shows significant variations after the profile change, connected to complex changes to the profile structure, including features that appear, decay or sustain over time (Figures \ref{fig:waterfalls_Eph} and \ref{fig:waterfalls_FFT}). These changes have a substantial, positive dependence on radio frequency (Figure \ref{PCA_models}). Similarly, while no significant variability is evident in the circularly polarised profiles at lower frequencies, the profile change event has introduced mild variations in Stokes $V$ above $\sim 2$\,GHz (Section \ref{ProVar}), inconsistent with cold-plasma propagation effects \citep{2016ApJ...817...16C}.

While the magnitude of the profile change has greatly diminished, it is unclear whether the profile is tracking toward full recovery or has permanently reconfigured; our phenomenological models point to either option for different frequency sub-bands. While the results for Stokes $I$ predominantly point towards an ongoing push towards a full recovery, our results for linear polarisation show mixed support for a permanent reconfiguration. Ongoing long term monitoring will resolve this ambiguity.

The April 2021 PSR~J1713$+$0747 profile change event, the largest and longest-lasting observed profile change event on any MSP, may provide valuable insights into these complex objects. While stochastic profile variability is common in the canonical pulsar population, discrete profile change events -- whilst rare -- have been observed in the MSP population. However, as datasets grow and telescope sensitivities improve over time, we may begin to detect profile changes in other MSPs, albeit on smaller scales. Understanding MSP profile changes as we head into the SKA era, their recoveries, and their impacts on ongoing precision timing experiments, such as those conducted with PTAs, is crucial for the continued study of these phenomena and the objects themselves.

\begin{acknowledgement}
We thank Simon Johnston and Lucy Oswald for thoughtful discussions and suggestions which improved this work. 
Murriyang, CSIRO’s Parkes radio telescope, is part of the Australia Telescope National Facility (\url{https://ror.org/05qajvd42}) which is funded by the Australian Government for operation as a National Facility managed by CSIRO. We acknowledge the Wiradjuri people as the Traditional Owners of the Observatory site. We acknowledge the Wallumattagal people as the Traditional Owners of the land where this work was carried out. This paper includes archived data obtained through the CSIRO Data Access Portal (http://data.csiro.au). Work at NRL is supported by NASA. MEL is supported by an Australian Research Council (ARC) Discovery Early Career Research Award DE250100508. Parts of this research were conducted by the Australian Research Council Centre of Excellence for Gravitational Wave Discovery (OzGrav), project number CE230100016.

Software: PSRCHIVE \citep{2010PASA...27..104V,2011ascl.soft05014V,2012AR&T....9..237V}, TEMPO2 \citep{2006MNRAS.372.1549E,2006MNRAS.369..655H}\\

\textbf{Profiles, profile residuals, and analysis scripts used in this paper are available at \url{https://github.com/CosmicRami/PSRJ1713-0747-Profile-Change}.} 

\end{acknowledgement}



\end{document}